\documentclass[%
 reprint,
 amsmath,amssymb,
 aps,
]{revtex4-2}
\usepackage{xcolor}
\usepackage{comment}
\usepackage{graphicx}
\usepackage{dcolumn}
\usepackage{bm}
\usepackage{xcolor}

\begin{document}

\preprint{APS/123-QED}

\title{Dynamics of Charge-Density-Wave \textit{puddles} in 2\textit{H}-NbSe$_2$ 
}

\author{Shreya Kumbhakar$^{1,2,\star}$}
\author{Marina Esposito$^{2,3,\star}$}
\author{Anjan Kumar N M$^{1,\star}$}
\author{Liwen Feng$^{1}$}
\author{Tommaso Confalone$^{2,4}$}
\author{Vasilisa Gerega$^{2,4}$}
\author{Flavia Lo Sardo$^{2,5}$}
\author{Rafiqul Alam$^{2}$}
\author{Davide Masarotti$^{6}$}

\author{Francesco Tafuri$^{3}$}
\author{Thomas Böhm$^{2}$}
\author{Mahmoud Abdel-Hafiez$^{7}$}
\author{Sushmita Chandra$^{8}$}
\author{Claudia Felser$^{8}$}
\author{Kornelius Nielsch$^{2,5}$}

\author{Nicola Poccia$^{2,3,\dagger}$}
\author{Stefan Kaiser$^{1,\dagger}$}
\author{Golam Haider$^{2,\dagger}$}

\affiliation{$^1$Institute of Solid State and Materials Physics, University of Technology, Dresden, Germany}
\affiliation{$^2$Leibniz Institute for Solid State and Materials Research Dresden (IFW
Dresden)
Dresden 01069, Germany}
\affiliation{$^3$Department of Physics, University of Naples Federico II, Naples 80126, Italy}
\affiliation{$^4$Institute of Applied Physics
Technische Universität Dresden
01062 Dresden, Germany}
\affiliation{$^5$Institute of Materials Science
Technische Universität Dresden
01062 Dresden, Germany}
\affiliation{$^6$Department of Electrical Engineering and Information Technology, University of Naples Federico II, Naples I-80126, Italy}
\affiliation{$^7$Department of Applied Physics and Astronomy, University of Sharjah, United Arab Emirates}

\affiliation{$^8$Max Planck Institute for Chemical Physics of Solids, 01187 Dresden, Germany}
\affiliation{$\star$ equal contribution}
\affiliation{$\dagger$ g.haider@ifw-dresden.de, stefan.kaiser@tu-dresden.de, nicola.poccia@unina.it}
\begin{abstract}
Electronic phases in quantum materials are often governed by nanoscale inhomogeneity, where local order develops within spatially confined regions or \textit{puddles}. A prominent example is an incommensurate charge-density-wave (I-CDW) that comprises locally commensurate domains. In 2\textit{H}-NbSe$_2$, such an I-CDW state persists alongside lattice anharmonicity and superconductivity, raising fundamental questions about the dynamical stabilization of CDW order in puddles. Here, we probe the puddle-dynamics in 2\textit{H}-NbSe$_2$. Raman scattering reveals a strong Fano-coupling between the interlayer shear phonons and the CDW amplitude mode. Time-resolved reflectivity measurement shows a low-frequency $\sim0.2$~THz coherent overdamped oscillation onsetting within the CDW regime at $\sim17$~K, pointing towards a so far unexplored transition. This we identify as a Fano-coupled phonon-CDW hybrid emerging from the collective dynamics of CDW puddles. \textcolor{black}{We establish a framework for capturing intertwinned glassy dynamics with an overdamped oscillatory model.} Such dynamics in 2H-NbSe$_2$ highlight how lattice pinning and electronic correlations in layered materials affect the CDW order, which is crucial for the design of novel Van der Waals devices.
\end{abstract}

\maketitle

Strongly correlated electron systems, in which lattice, charge, spin, and orbital degrees of freedom interact actively with each other, naturally develop local inhomogeneities or `\textit{puddles}' of distinct types of order~\cite{doi:10.1126/science.1107559,doi:10.1126/science.1150124,dagotto2002nanoscale,PhysRevB.84.060511,joe2014emergence,sajan2025atomic,PhysRevLett.132.056401,PhysRevLett.76.3412,bianconi1996stripe,poccia2012optimum,fratini2010scale,poccia2012optimum,campi2015inhomogeneity}, driven by the competition of different phases~\cite{emery1990phase,kivelson1998electronic}.
Recent experiments~\cite{chu2023fano,PhysRevB.84.020509,caivano2008feshbach} suggest that 
the coexistence of these puddles can generate a multiband or multigap system~\cite{PhysRevB.84.020509,homeier2025feshbach}, where new interference effects emerge from a generalized `Fano' coupling~\cite{Joe_2006} between the different bands. 
Importantly, such systems, ranging from ultracold gases to high-Tc cuprates~\cite{PhysRev.124.1866,FESHBACH1958357,chin2010feshbach,miroshnichenko2010fano,bianconi2013shape,bianconi2000stripe,innocenti2010resonant}, can be tuned close to a Fano resonance condition through doping, external pressure, internal chemical pressure or `microstrain' in layered materials, magnetic fields, or geometrical control~\cite{PhysRevB.84.020509,caivano2008feshbach,kugel2008model,poccia2009misfit,PhysRevB.110.184510}, where particle–particle correlations and scattering intensities are strongly enhanced.
Here, we investigate 2\textit{H}-NbSe$_2$, a prominent system where both superconductivity (SC) and unconventional CDW phases coexist at low temperatures~\cite{kiss2007charge,PhysRevLett.102.166402,PhysRevResearch.2.043392,sanna2022real,PhysRevB.94.140504,PhysRevB.108.L100504,PhysRevB.108.235160,PhysRevB.89.060503}. 
The CDW phase comprises puddles of distinct local CDW commensurations \cite{{pasztor2021multiband},PhysRevB.89.235115,PhysRevLett.132.056401}.
We study the temporal dynamics of the puddles, which remain elusive despite extensive investigation of their spatial distribution~\cite{soumyanarayanan2013quantum, PhysRevB.89.235115,PhysRevB.89.235115,{pasztor2021multiband},PhysRevB.89.235115,PhysRevLett.132.056401,sheng2024terahertz}. The simpler chemical structure of NbSe$_2$ among materials with complexity~\cite{dagotto2002nanoscale} helps in disentangling the CDW-puddle dynamics from other intertwined orders~\cite{rohwer2011collapse,PhysRevB.108.L100504,doi:10.1126/sciadv.aax3346,chu2023fano}, making it a suitable material choice. 
\begin{figure}[ht!]
    \centering
  
\makeatletter
\renewcommand{\fnum@figure}{\textcolor{black}{\figurename~\thefigure}}
\makeatother
\includegraphics[width=0.37\textwidth]{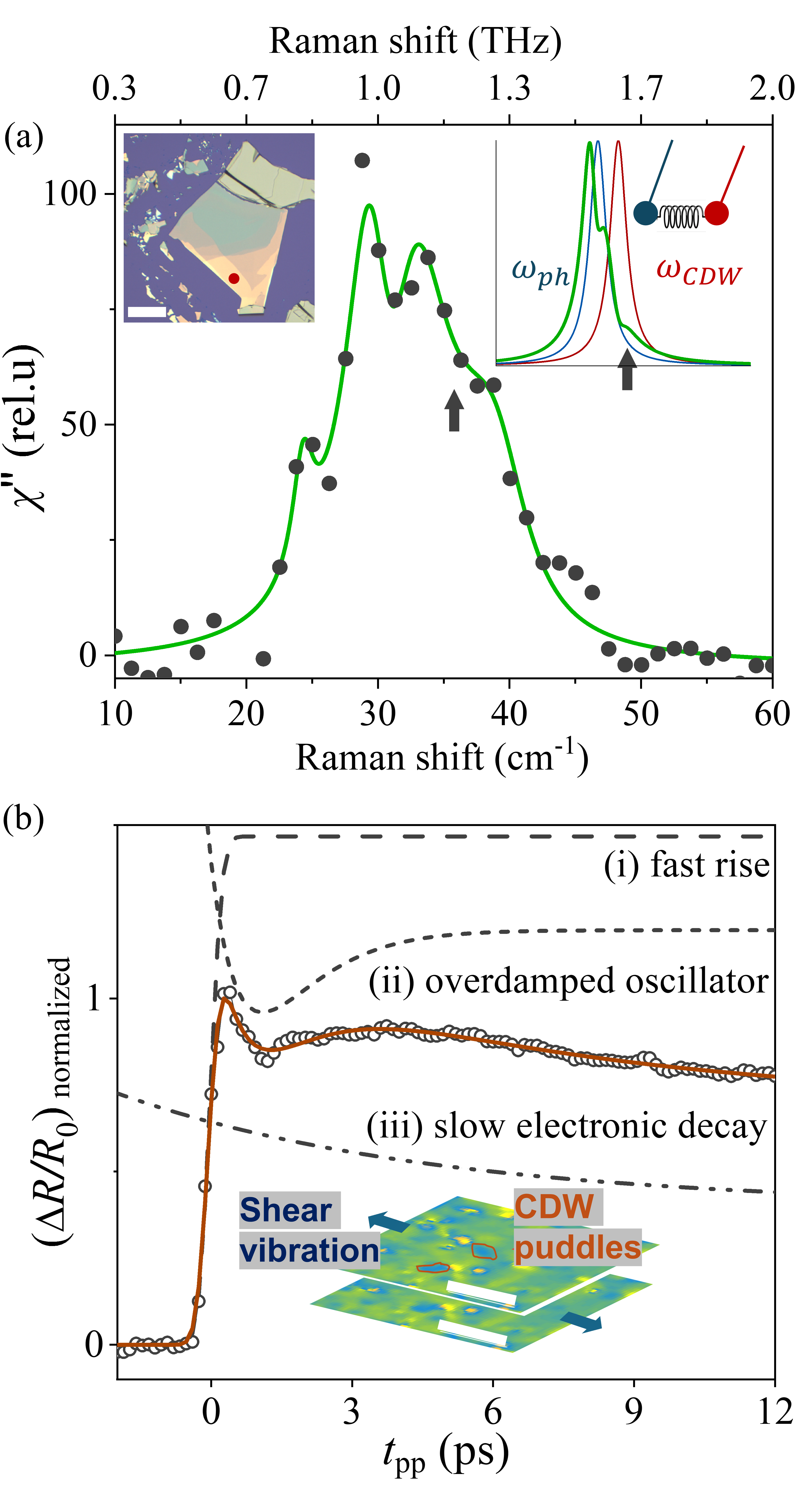} 
    \caption{\textbf{Fano-coupled phonon-CDW response.}
(a) Raman susceptibility, $\chi''$ measured in unpolarized backscattering configuration on a mechanically exfoliated bulk-flake (thickness $\sim60$~nm) of NbSe$_2$ on Si/SiO$_2$ substrate at 14~K is shown. The arrow points to a broad shoulder at $\sim40$~cm$^{-1}$. The green solid line shows a fit to the data using a Fano resonance model, as sketched in the right inset of Fig.~1a.
Left inset: Optical image of the flake of NbSe$_2$, with the red dot pointing the point of Raman measurements.
Right inset: Fano coupling between two discrete oscillators, modelled by Lorentzian lineshapes (blue and red solid curves), leads to the Breit-Wigner Fano lineshape (green solid line). The black arrow represents the shoulder, which is analogous to the data in (a).
(b) Time-resolved reflectivity change ($\Delta R / R_0$), normalized with respect to its peak value, of NbSe$_2$ single crystals measured at $T\sim2$~K with \textcolor{black}{80~$\mu$J/cm$^2$} of 1.2 eV pump fluence is shown. \textcolor{black}{$R_0$ is the static probe reflectivity and $t_{pp}$ is the pump-probe delay}. The data is fitted by a convolution of three distinct components, which are shown separately by the dashed, dotted, and dashed-dotted lines, respectively and discussed further in the main text.
(Inset:) The schematic depicts CDW puddles (represented by regions marked with red solid lines). \textcolor{black}{The scale bar corresponds to 10~nm length scales.} The blue arrows indicate the interlayer shear vibration that couples with the CDW. The colour maps in the square sheets, indicating the layers, are CDW modulations measured via STM experiments reproduced from Ref.~\cite {pasztor2021multiband}. 
}
\label{fig:1} 
\end{figure}

The puddle-dynamics can provide unique insights into the formation of the CDW order itself.
For example, although phonons soften strongly near the CDW wave vector in NbSe$_2$~\cite{PhysRevB.94.140504,PhysRevB.92.140303}, it remains debated whether phonons alone drive the transition or whether electronic correlations play an essential role. 
One of the primary anomalies causing this debate is the persistence of CDW-like spectral features above $T_{\mathrm{CDW}}$ before the onset of phonon-softening or in the absence of substantial lattice distortions.
The microscopic character of the pronounced short-range CDW order observed well above $T_{\mathrm{CDW}}$~\cite{PhysRevLett.102.166402, chatterjee2015emergence,soumyanarayanan2013quantum, PhysRevB.89.235115} remains unclear \textit{i.e.} whether they represent amplitude vs. phase incoherence, domain dynamics, or preformed electronic order.
The very nature of CDW in NbSe$_2$ is also not understood clearly as to why the CDW wave vector does not coincide with a strong nesting feature of the Fermi surface~\cite{PhysRevB.77.165135, flicker2015charge} and is not perfectly commensurate, raising the question of how anharmonic lattice effects \cite{PhysRevB.94.140504,PhysRevB.92.140303}, momentum- and orbital-dependent electron-phonon coupling~\cite{flicker2015charge}, and higher-order free-energy terms stabilize a particular triangular pattern of the CDW domains \cite{cao2024directly,PhysRevLett.132.056401,doi:10.1021/acs.nanolett.8b00237}.

In this study, we perform Raman scattering and ultrafast spectroscopy measurements on bulk-like mechanically exfoliated flakes of 2\textit{H}-NbSe$_2$. We find that the  CDW amplitude mode exhibits a strong Fano lineshape that captures the coupling between the interlayer shear mode and the CDW mode. 
Time-resolved reflectivity of the crystals reveals an overdamped oscillation that is intertwined with the electronic response, tracking both CDW and SC orders, setting in at $T_{\mathrm{CDW}}\sim 28$~K and $T_c\sim 7$~K, respectively. 
The divergence of the electron relaxation time in the ultrafast response, together with the low-frequency of the oscillation at $\sim14-17$~K within the CDW phase, establishes an onset of a coherent collective dynamics of the CDW-puddles, which is \textit{glassy} in nature.
A correlation between the Fano coupling parameter and the oscillation frequency suggests that a competition between different CDW commensuration orders and an early onset of SC fluctuations is reflected in the puddle dynamics.

Figure~1(a) shows the unpolarized Raman susceptibility, $\chi^"$, measured in bulk flakes of NbSe$_2$ mechanically exfoliated inside a glovebox and loaded in-situ for environment-protected Raman measurements. (See left inset of Fig.~1a and {Materials and Methods} for more details). We focus on the low-frequency spectral region up to 60 cm$^{-1}$. 
We observe a peak at $29~$cm$^{-1}$, corresponding to the interlayer shear vibration~\cite{He_2016} and another peak at $35~$cm$^{-1}$, at which the CDW amplitude mode~\cite{PhysRevLett.37.1407,PhysRevLett.45.660,PhysRevB.89.060503,PhysRevLett.134.066002} has been reported. The arrow in the figure indicates a broad shoulder at $40$~cm$^{-1}$. There is also a prominent notch at 25~cm$^{-1}$. 
These features reveal a coupling between the interlayer phonon and the CDW order that is accurately described by a Fano lineshape (green solid line)~\cite{cardona2005light, PhysRevB.20.5157, blumberg1994investigation}, sketched in the right inset of Fig.~1(a). (See {Materials and Methods} for discussion.)
Such a Fano coupling is intimately connected to the presence of puddles of local order in the  system~\cite{chu2023fano, PhysRevB.84.020509,caivano2008feshbach,PhysRevB.84.020509,homeier2025feshbach}, in this case, the CDW puddles~\cite{pasztor2021multiband,PhysRevB.89.235115, PhysRevLett.132.056401}, motivating us to investigate its dynamics.  
Fig.~1(b) shows the time-resolved ultrafast optical reflectivity changes, $\Delta R/R_0$
at 2~K in single crystals of NbSe$_2$. (See {Materials and Methods} for details on the pump-probe experiments.) We observe and accurately model the response as a convolution of (i) a fast rise in $\Delta R/R_0$ close to zero pump-probe delay ($t_{pp}=0$), followed by (ii) an overdamped oscillation before approaching (iii) a slow decay (See {Materials and Methods} for discussion on the fitting.) 
We find this overdamped oscillation has a low-frequency $\sim 0.18$ THz, compared to the CDW amplitude mode $\sim1.2$~THz~\cite{PhysRevLett.45.660,sheng2024terahertz,PhysRevB.89.060503,PhysRevB.107.245125}. In the following, we are going to identify this as a new collective mode in the system emerging from the CDW puddles.
\textcolor{black}{The characteristic puddle dimensions reported by STM are typically $\sim10$~nm~\cite{soumyanarayanan2013quantum, PhysRevB.89.235115,PhysRevB.89.235115,{pasztor2021multiband},PhysRevB.89.235115,PhysRevLett.132.056401,sheng2024terahertz}, far below the optical spot sizes employed in the present Raman and ultrafast experiments. Our measurements therefore probe the collective dynamics driven by a macroscopic coherent ensemble of puddles.}
\begin{figure}[ht!]
    \centering
  \makeatletter
\renewcommand{\fnum@figure}{\textcolor{black}{\figurename~\thefigure}}
\makeatother
\includegraphics[width=0.48\textwidth]{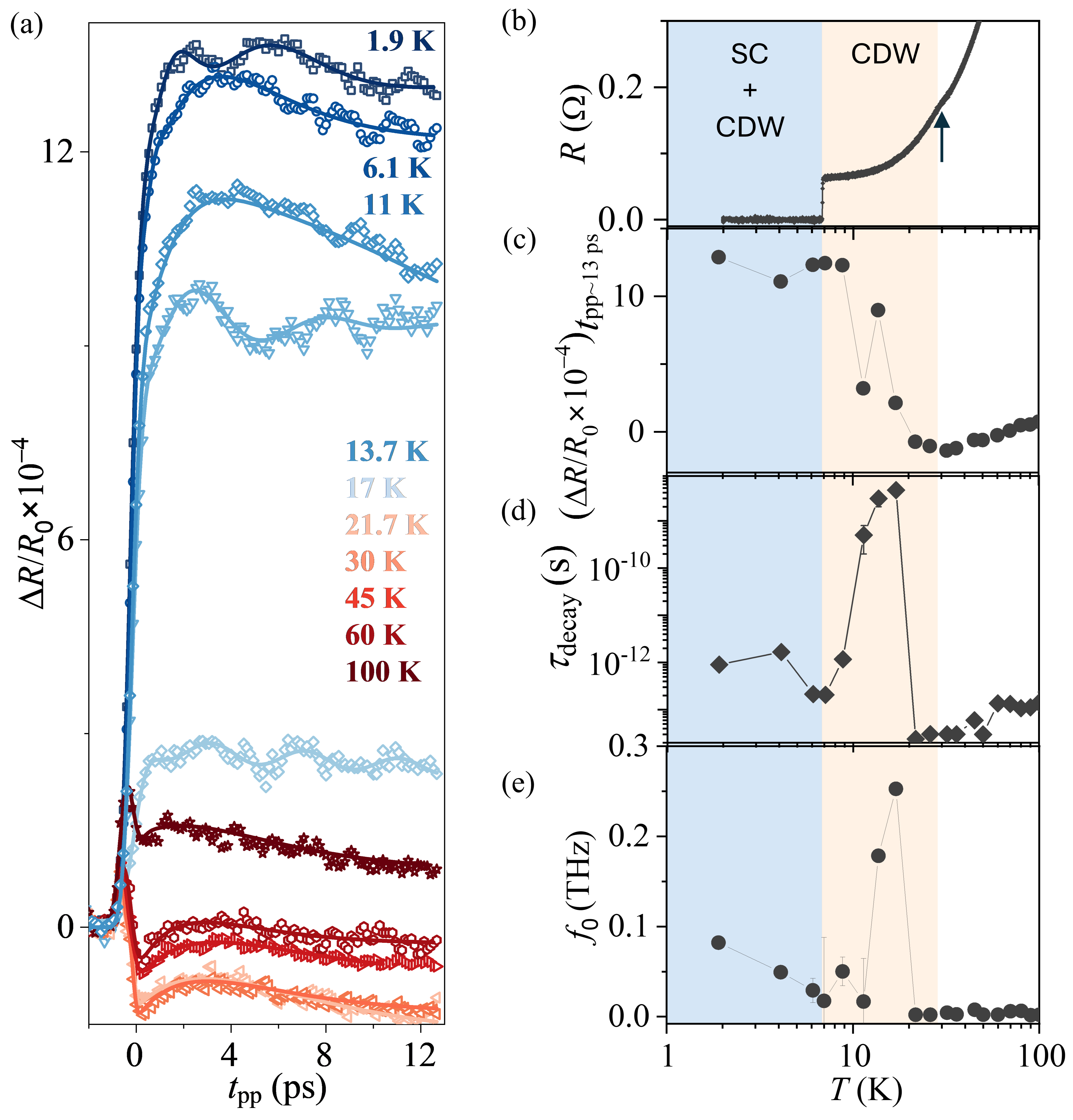} 
    \caption{\textbf{Temperature-dependent ultrafast dynamics in bulk NbSe$_2$:} (a) The time-resolved reflectivity, $\Delta R/R_0$ is shown at different temperatures ($T$s) from 1.9~K-100~K. The open points are the measured data and the solid lines are the respective fits using the model discussed in Fig.~1(b). (b) Representative $T$-dependent four-probe resistance, $R$ of the NbSe$_2$ crystals is shown. $R$ jumps to nearly zero resistance ($\approx1$~n$\Omega$) at $T_c\sim7$~K, indicating the superconducting transition. The data shows a kink at the CDW transition temperature, $T_{\mathrm{CDW}}\sim28$~K, marked by the solid arrow. The blue and beige regions are marked based on these transition temperatures. (c)-(e) show the $T$-dependence of the long-lived reflectivity, \textit{i.e.} $\Delta R/R_0$ at $t_{pp}\sim13$~ps, 
    the relaxation time, $\tau_{\mathrm{decay}}$, and overdamped oscillation frequency, $f_0$, estimated from the fits, respectively.}
    \label{fig2}
\end{figure}

In Fig.~2(a), we show the $T$-dependent ultrafast response in NbSe$_2$ from 100~K down to 2~K (see Supporting Information {SI Sec II} for detailed $T$-dependence). 
We observe a noticeable drop in the overall amplitude of the $\Delta R/R_0$ above $T_c\sim6-7$~K, 
 and even more significant reduction above $\sim13.7$~K. Furthermore, above 17~K, $\Delta R/R_0$ changes sign. We will show in further analysis that 17 K is also the temperature below which the overdamped oscillation (Fig.~1(b)) sets in. \textcolor{black}{This we are going to identify as the onset of a glassy dynamics.}
The largest negative response in $\Delta R/R_0$ is reached close to $T_{\mathrm{CDW}}\sim28$~K. 
The transition temperatures for SC and CDW, $T_c$ and $T_{\mathrm{CDW}}$, respectively, have been extracted from the transport characterization in Fig~2(b), with blue and beige-shaded regions marking the corresponding phases in Fig.~2(b)-2(e).
These changes in the ultrafast response have been quantitatively captured by the long-lived reflectivity change, $\Delta R/R_0$ at $t_{pp}\sim13$~ps, as shown in Fig.~2(c).
\textcolor{black}{Fig. 2a also shows the overdamped oscillations being more prominent at the lowest $T$s below $T_c$ and again close to 17 K, with distinct oscillation frequencies.}
Analyzing $\Delta R/R_0$ of Fig.~2(a) with the model discussed in Fig.~1(b) (See Eq.~[2] in Materials and Methods), Fig~2(d), and (e) show the oscillation frequency ($f_0$), and the decay time ($\tau_{decay}$), respectively. \textcolor{black}{Note that a conventional multichannel exponential decay model fails to capture the oscillatory features of the glassy dynamics. See SI Sec I for more details.}

$\tau_{\mathrm{decay}}$ (in Fig.~2(d)) shows a significant increase first at 14 K, and then another rise at $T_c$. 
The divergence of $\tau_{\mathrm{decay}}$ is generally connected to a quasiparticle recombination bottleneck effect  (e.g., Rothwarf–Taylor bottleneck) associated with gap opening, or from critical slowing down of collective order parameter dynamics near the phase transition~\cite{Giannetti03032016,yusupov2010coherent}. The increase in $\tau_{\mathrm{decay}}$ close to $T_c$ is consistent with the opening of a SC gap~\cite{PhysRevB.59.1497,PhysRevLett.83.800,Giannetti03032016}. However, the significant enhancement of $\tau_{\mathrm{decay}}$ at 14-17~K has not been discussed before. Interestingly, $\tau_{\mathrm{decay}}$ has no sharp change at $T_{\mathrm{CDW}}$, similar to previous report~\cite{{PhysRevB.108.235160}}. This is probably because the CDW gap has been observed to be present until room $T$, though incoherent in nature~\cite{chatterjee2015emergence}, with the coherence setting in only at $T_{\mathrm{CDW}}$~\cite{PhysRevLett.102.166402}. 
\textcolor{black}{On the other hand, the diverging behavior of $\tau_{\mathrm{decay}}$ at $14-17$~K, where no electronic gap opening has been reported, implies a critical slowing down of electrons from the onset of a glassy dynamics.
}\textcolor{black}{These values essentially span the hierarchy of relaxation scales, which has been established independently by Anikin et al. \cite{PhysRevB.102.205139} using broadband pump–probe spectroscopy of 2H-NbSe$_2$. Their global analysis involving a multi-exponential decay model resolves characteristic times of approximately 100 fs, 1 ps, 100 ps and 1 ns. These are associated, respectively, with electron–electron thermalization, electron–phonon relaxation, anharmonic phonon–phonon relaxation and heat dissipation. We note that in the convoluted dynamics, $\tau_{deacy}$ should not be interpreted as the lifetime of one fixed microscopic channel over the entire $T$-range, but the dominant relaxation pathway of the itinerant electronic subsystem. (For details See Sec I). Here, the change at 17 K indicates the crossover from a fast IRF-limited electron-electron channel to thermal dissipation and an anharmonic phonon-phonon channel due to the onset of coupled lattice-CDW dynamics.}

\textcolor{black}{We observe $f_0$ (in Fig.~2(e)) to be nearly zero above 17 K. At 17 K, it increases sharply to $\sim 0.25$ THz.}
Below 14~K, we again observe a softening of $f_0$, down to $\sim0.05$~THz. Its value remains almost constant until 7~K, close to $T_c$, below which it shows a monotonic increase. 
The softening of $f_0$ to almost zero at higher $T$s indicates that the oscillation does not emerge from a simple phononic mode (like coherent phonons), but rather a \textit{new} collective order in the system, which in this case is the CDW order. The low-frequency of the oscillation ($\sim0.2-0.25$~THz) further indicates that this is not driven purely by the (fast) CDW amplitude mode (at $40~\mathrm{cm}^{-1}\simeq1.2~$THz, \textcolor{black}{as can also be seen from the Raman response in Fig.~1(a)}) but rather by the (slower) fluctuations of CDW puddles. \textcolor{black}{Also, a recent THz-STM study of 2H-NbSe$_2$ reported CDW-specific collective dynamics on a similar sub-THz frequency scale, including a mode near 0.15 THz, attributed to spatially heterogeneous, defect-pinned CDW phase dynamics~\cite{sheng2024terahertz}.}

\textcolor{black}{We make two key observations here. Firstly, the coherent oscillations set in at 17 K, in the same $T$-window as $\tau_{\mathrm{decay}}$ diverges, supporting 
the onset of a \textit{glassy} dynamics of CDW-puddles. Glassy dynamics often appear from electronic frustration and/or disorder~\cite{PhysRevB.79.165122}. In this case, we believe a strong competition between different CDW commensurations promotes this~\cite{PhysRevLett.85.836,ricci2021measurement,campi2022nanoscale}. Secondly, 
the frequency softens again at 14 K (to $\sim0.1$~THz) and increases monotonically below $T_c$. This softening may be related to the onset of SC fluctuations above $T_c$ that affect the puddle dynamics~\cite{joe2014emergence, PhysRevLett.134.066002,PhysRevX.13.021008,sachdev2000thermally}.
} 
 \begin{figure}[ht]
    \centering
\makeatletter
\renewcommand{\fnum@figure}{\textcolor{black}{\figurename~\thefigure}}
\makeatother\includegraphics[width=0.48\textwidth]{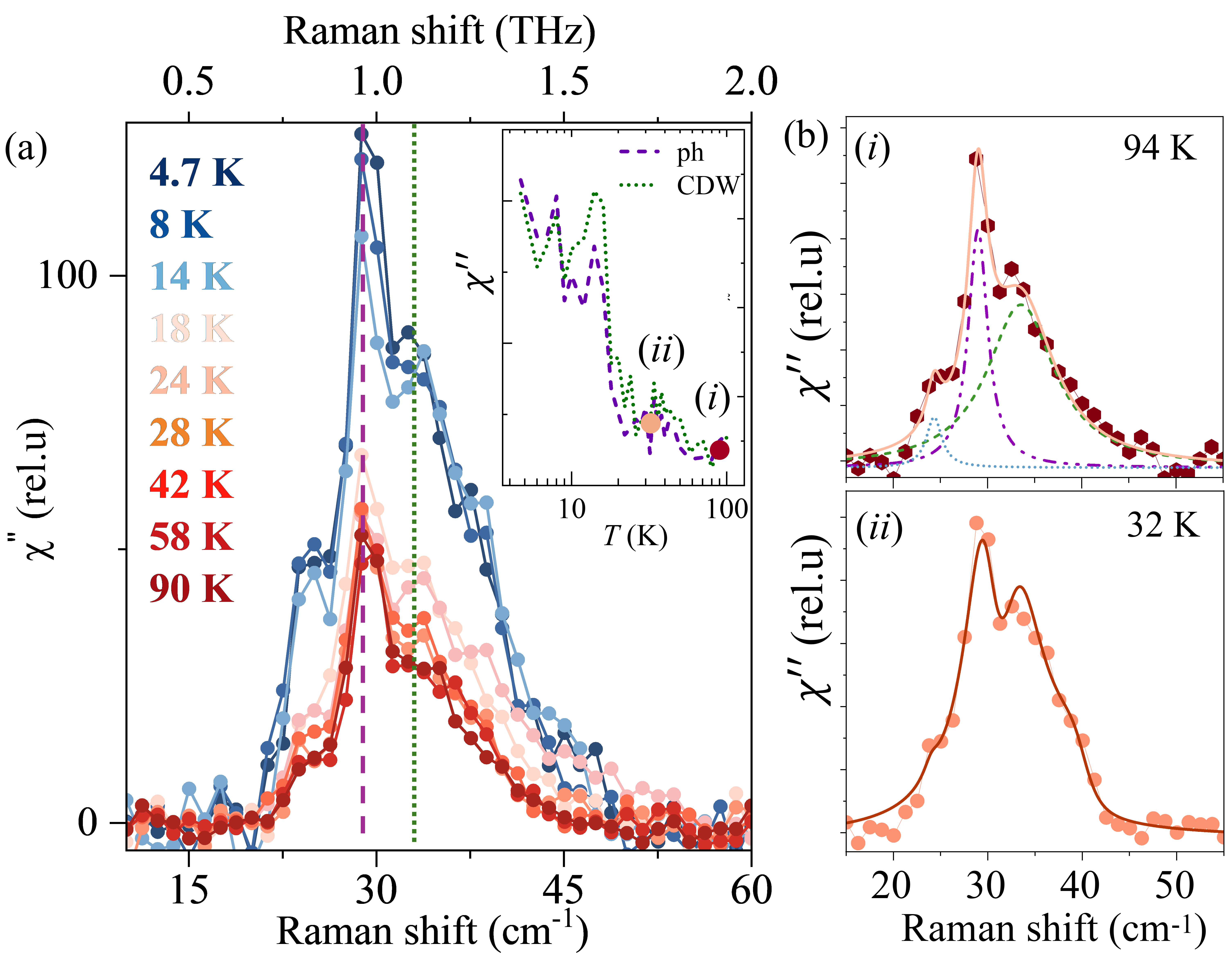}
    \caption{
    \textbf{Temperature dependence of the Raman susceptibility of bulk NbSe$_2$. }
    (a) Raman susceptibility ($\chi^"$) is measured at different $T$s on bulk flakes of NbSe$_2$. The inset plots the value of $\chi^"$ at the frequencies marked by the dashed and dotted lines. 
    (b) $\chi^"$ is plotted at two distinct $T\sim~32,~\mathrm{and}~94$~K. The solid lines represent fits to the data. At 94 K, the data is fitted with additive contributions from three distinct lorentzian lineshapes, depicted by the dahed lines, which represent the interlayer shear phonon peak at $\sim29$ cm$^{-1}$, CDW amplitude mode at $\sim35$ cm$^{-1}$, and another peak at 25~cm$^{-1}$, that is identified with a split E$_{2g}$ peak of the interlayer shear vibration. The data at 32~K is fitted by a Fano lineshape, as we discuss further in the text.}
    \label{fig:3}
\end{figure}
\begin{figure}[ht]
    \centering
  \makeatletter
\renewcommand{\fnum@figure}{\textcolor{black}{\figurename~\thefigure}}
\makeatother
\includegraphics[width=0.45\textwidth]{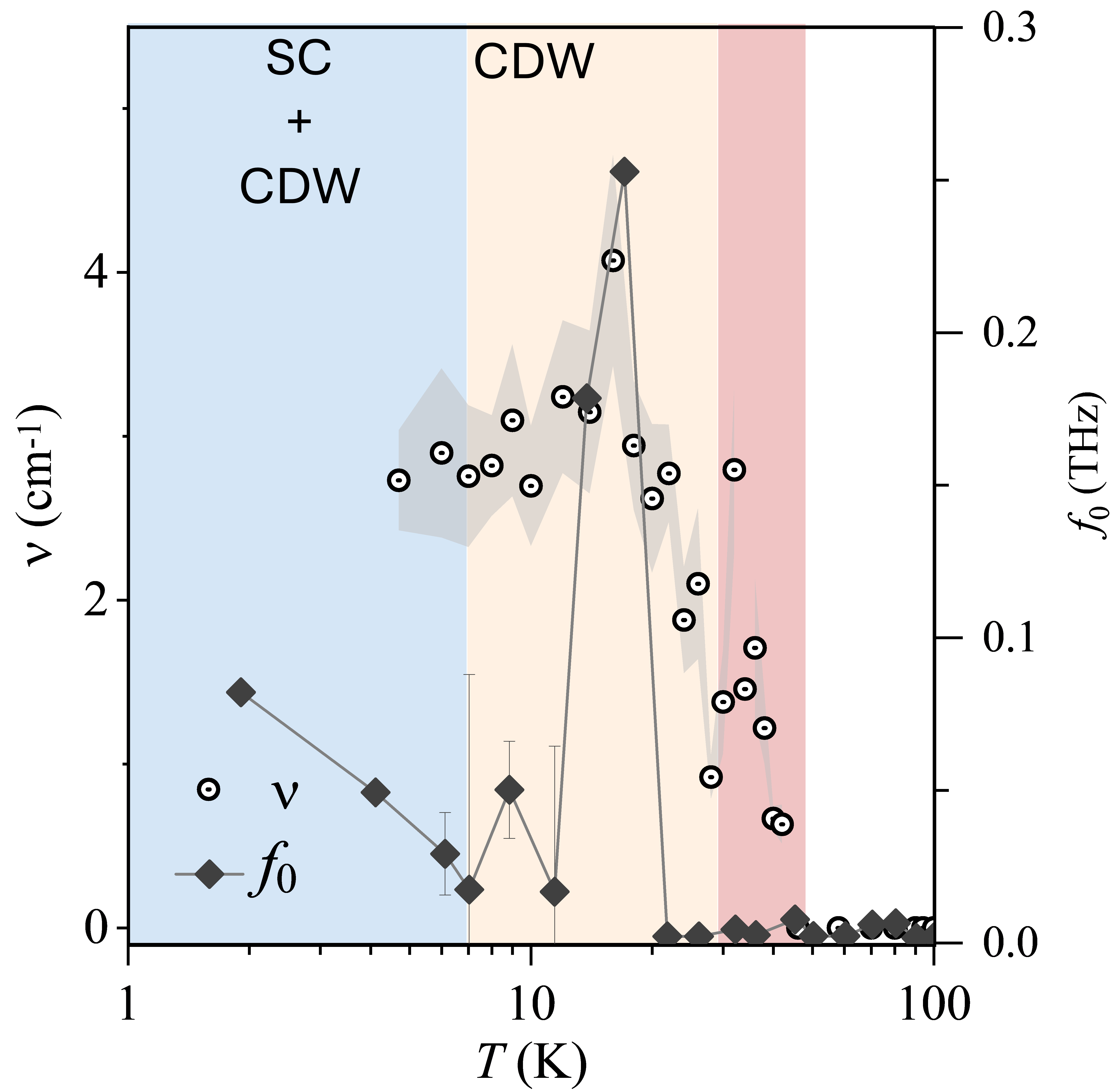} 
    \caption{\textbf{Temperature-dependence of the Fano coupling factor} $\nu$, estimated from the fits of the Raman susceptibility, $\chi^"$ at different $T$s is plotted on the left-axis. The grey-shaded region represents the fitting error. The right-axis plots the overdamped oscillation frequency, $f_0$, obtained from the ultrafast response. Blue and beige shaded regions are marked based on the superconducting and CDW transition temperatures, characterized from electrical transport in Fig.~2(a). The red region marks, the onset of $\nu$ as discussed further in the text.}
    \label{fig:4}
\end{figure}

To further characterize this complex interplay of phases, we have carried out $T$-dependent Raman measurements, as shown in Fig.~3(a). (See {SI Sec IV} for detailed $T$-dependence.) 
First, we observe a noticeable reduction in the overall intensity of the spectra above 17~K. This is quantitatively captured by the $T$-dependence of $\chi^"$ at $29$~cm$^{-1}$ and $35$~cm$^{-1}$ (associated with the interlayer shear vibration and CDW amplitude modes respectively) shown in the inset. 
Second, the features at 35 cm$^{-1}$ and 40 cm$^{-1}$, which we attribute to a ph-CDW Fano coupling effect (discussed in Fig.~1(a)), have non-trivial $T$-dependences. Although a peak at 35 cm$^{-1}$ can be identified until 100 K, the broad shoulder centred at 40 cm$^{-1}$, can be clearly resolved only at lower temperatures (below $T\sim50$ K as we show later). 
We also note from the inset nearly constant magnitudes of $\chi"$ with decreasing $T$ below $T\sim 50$~K until $17$~K, indicating changes in the Fano coupling behaviour at these temperatures.
To better understand this, we plot the Raman spectra at two representative temperatures, $T\sim94$~K and 32~K in the top and bottom panels of Fig.~3(b).
At $T\sim94$~K, the contribution from three distinct Lorentzians captures the data accurately. 
We note the persistence of the peak at $35$~cm$^{-1}$ above $T_{\mathrm{CDW}}$ is consistent with the observation of the CDW gap even until higher $T$s~\cite{chatterjee2015emergence}. 
Furthermore, we believe that the peak at 25~cm$^{-1}$ is a splitting (of $\approx 4$~~cm$^{-1}$) from the doubly degenerate E$_{2g}$ peak at 29~cm$^{-1}$ of the interlayer shear vibration due to strain induced from CDW fluctuations and is hence intimately tied to the CDW amplitude mode. Though a splitting of E$_{2g}$ peaks from strain has been experimentally and theoretically measured in intercalated NbSe$_2$ layers from strain~\cite{PhysRevB.110.075430}, the same has not been observed on bulk 2\textit{H}-NbSe$_2$ layers.
At lower temperatures (e.g., at 32 K in the bottom panel of Fig.~3(b)), we clearly observe the presence of a broad shoulder at 40 cm$^{-1}$, 
that is accurately modelled by a Fano resonance between the phonon and CDW amplitude modes 
({See Materials and Methods for fit details}). 
Figure~4 plots the $T$-dependence of the Fano coupling parameter, $\nu$, extracted from the Raman response along with $f_0$. 
We find a strong fluctuating onset below 50 K, that starts showing a monotonic increase below $T_{\mathrm{CDW}}\sim28$~K. 
At $17$~K, $\nu$ reaches a maximum value, below which it decreases again, yielding a stable plateau from $\sim14$~K onwards to the lowest measurable temperatures.
We note that these temperatures do not manifest in the phase characterization from transport (shown by blue and beige regions).
Thus, our data capture transitions marking the onset of novel dynamics. (See {SI Sections I, V, VI, VII} for additional experimental support.)  
The Fano coupling is intimately related to phase-seperation or phase-coexistence in a system~\cite{PhysRevB.84.020509,homeier2025feshbach,PhysRevB.84.020509,caivano2008feshbach}, which in this case is the CDW puddles. 
\textcolor{black}{From the onset of $\nu$ below 50~K, we conclude an \textit{incoherent} puddle regime begins at this temperature, that is well-above T$_{\mathrm{CDW}}$, marked by the red shaded region. A collective coherent motion only sets in at 17 K, captured by the increase of $f_0$ where $\nu$ attains its maximum strength likely from the competing CDW orders. This indicates that local response from the puddles, represented by $\nu$, captures the emerging glassy collective dynamics at $\sim17$~K.
With further decrease in $T$, $\nu$ stabilizes below $\sim 14$~K where $f_0$ decreases as well. We could interpret this as a possible onset of SC above $T_c$ (as also proposed in \cite{PhysRevLett.134.066002}), that impacts the CDW order and commensuration~\cite{doi:10.1021/acs.nanolett.8b00237,PhysRevB.108.L100504} and is thus manifested in the puddle dynamics.}
In summary, both ultrafast and Raman measurements reveal the onset of a new dynamics in 2\textit{H}-NbSe$_2$ that is driven by the local CDW domains, or {puddles}.
Locally, this manifests as a Fano coupling between the interlayer shear vibration and CDW amplitude mode in the Raman spectra, which directly corroborates the dynamics measured in the ultrafast response. Such a coupling sheds insight into multiple longstanding questions about the CDW phase in NbSe$_2$. First, it explains why CDW amplitude mode oscillations corresponding to $40$~cm$^{-1}$ could not be clearly observed in the ultrafast response before.
Second, it suggests an enhanced coherence of the CDW phase with decreasing interlayer shear strain, providing a possible route for significant enhancement in $T_{\mathrm{CDW}}$ from the bulk to the monolayer~\cite{xi2015strongly}. Third, the Fano coupling can aid in CDW incommensuration in NbSe$_2$, which remains an open question~\cite{PhysRevB.77.165135, flicker2015charge,PhysRevB.94.140504,PhysRevB.92.140303}. Fourth, the microscopic fluctuations observed above T$_{\mathrm{CDW}}$ \cite{PhysRevLett.102.166402, chatterjee2015emergence, soumyanarayanan2013quantum, PhysRevB.89.235115} likely represent preformed electronic order and phase incoherence, as we see signatures of the CDW amplitude mode even until 100 K.
Our results establish how local dynamics in puddles shapes the global response in a material. This can be used as a recipe to engineer novel functionalities in quantum materials.
\section*{Material and Methods}
\subsection*{Material growth}
High-quality single crystals of 2\textit{H}-NbSe$_2$ were grown using the conventional chemical vapour transport method. Initially, stoichiometric amounts of niobium powder (Alfa Aesar, 99.99$\%$) and selenium powder (Sigma-Aldrich, 99.98$\%$) were thoroughly mixed and sealed under vacuum in a quartz ampoule with iodine ($\sim3$~mg.cm$^{-3}$) as the transport agent. The sealed ampoule was placed in a dual-zone furnace with a temperature gradient of 800$^{\circ}$~C (source zone) and 720$^{\circ}$~C (growth zone) and maintained for a period of 15 days. After slowly cooling to room temperature, shiny layered NbSe$_2$ single crystals with typical lateral dimensions of a few millimeters were obtained.
\subsection{Raman Measurements}
Raman measurements were carried out in a polarized backscattering configuration using a WITec Alpha 300 confocal Raman microscope equipped with a $50\times$ objective (NA = 0.9), enabling a spot size of 1 micron with a continuous-wave 532 nm laser. 

High spectral resolution was achieved using a diffraction grating with 1800 lines mm$^{-1}$, providing a resolution of approximately 1.2 cm$^{-1}$. A 532 nm RayShield filter enabled the detection of low-frequency Raman modes down to 10 cm$^{-1}$ on both the Stokes and anti-Stokes sides of the spectrum.
The laser power at the sample was monitored with a calibrated power meter and set to 1 mW for single-point spectra. This power was confirmed not to ensure that the average heating is less than 1 K (60 s per spectrum, averaging over 10 accumulations). \textcolor{black}{See Sec IV anv V in SI for $T$-estimation and $T$-dependence of the high frequency phonon peaks to assess laser-induced heating effects.} 
Cryogenic measurements from 300 K to 4 K were carried out using a continuous flow cryostat Microstat HiRes from Oxford Instruments.

NbSe$_2$ flakes were exfoliated from the single crystals, grown as discussed above, on cleaned Si/SiO$_2$ substrates inside an Argon-filled glovebox and loaded in-situ in the cryostat for Raman measurements. This ensured pristine surfaces of the flakes were protected from environmental degradation. \textcolor{black}{For the results presented in this manuscript, we have focused on bulk flakes (thickness $\sim$ 60 nm comprising about 100 atomic layers), ruling out substrate-related effects. We also show the electrical transport characteristics of both the single crystals and exfoliated thick flakes in Sec I of SI.}
\subsection*{Ultrafast measurements}
We performed femtosecond nondegenerate pump–probe reflectivity measurements on high-quality single crystals. The ultrafast experiments used 190 fs laser pulses in a reflection setup, with a 1030 nm pump beam and a 680 nm probe beam. The laser pulses were generated by a regenerative amplifier system (Pharos, Light Conversion; 1030 nm central wavelength, 10 kHz repetition rate, 190 fs pulse duration). The output beam was split into two parts. One part was sent to an optical parametric amplifier to produce the 680 nm probe beam, while the 1030 nm beam was used as the pump to excite the sample.

The pump and probe beams were linearly polarized and set perpendicular to each other to reduce coherent artifacts and polarization-related interference effects. The time delay between the pump and probe pulses was controlled using a motorized delay stage. The pump beam was modulated with a mechanical chopper. The resulting pump-induced change in reflectivity ($\Delta R/R_0$) was measured using a photodiode and lock-in detection for improved signal sensitivity.

Temperature-dependent measurements were performed in a continuous-flow helium cryostat (Optistat CF, Oxford Instruments), which provided an inert environment and maintained a clean sample surface during the experiment. The full-width at half maximum (FWHM) spot sizes were about 200 $\mu$m for the pump and 100 $\mu$m for the probe, ensuring uniform excitation within the measured area. The pump and was kept constant at 40~$\mu$J/cm$^{2}$ (that is below the threshold  for the suppression of the CDW state~\cite{PhysRevB.102.205139,payne2020latticecontributionunconventionalcharge,PhysRevB.108.235160}), and 20~$\mu$J/cm$^{2}$, respectively, for all temperatures studied in Fig.~2.
\subsection*{Fano lineshape analysis of the Raman response}
The experimentally measured Raman spectra have been multiplied by the Bose factor~\cite{RevModPhys.79.175} to calculate the Raman susceptibility after subtracting a smooth background.
The spectra have been then fitted with a Fano lineshape~\cite{Klein1983, mialitsin2011fano, eklund1979analysis, blumberg1994investigation, Joe2006}, captured by the following expression:
   
\[
\begin{split}
f(\omega) =&{} \frac{A_{ph}^2\gamma_{CDW}^2 (\nu R(\omega) + \eta)^2}{(\omega - \omega_{CDW} - \nu^2 R(\omega) )^2 +(\nu^2 \pi r(\omega) + \gamma_{CDW})^2} \\
&+\frac{A_{ph}^2\pi r(\omega) [((\omega - \omega_{CDW} )+ \nu \eta)^2]}{(\omega - \omega_{CDW} - \nu^2 R(\omega) )^2 + (\nu^2 \pi r(\omega) + \gamma_{\mathrm{CDW}})^2}\\
&+\frac{A_{ph}^2\pi r(\omega)[\gamma_{CDW} (\nu^2 \pi r(\omega) + \gamma_{\mathrm{CDW}})]}{(\omega - \omega_{CDW} - \nu^2 R(\omega) )^2 + (\nu^2 \pi r(\omega) + \gamma_{\mathrm{CDW}})^2}
\end{split}
\]
where:
\[r(\omega)=\frac{\gamma_{ph}^2}{(\omega - \omega_{ph})^2 + \gamma_{ph}^2}\]
 and $R(\omega)$ is derived as Hilbert transform of $r(\omega)$.
Here, $\omega$ is the Raman-frequency shift. $\omega_{ph},\omega_{CDW}$ are the centre frequencies, $\gamma_{ph},\gamma_{CDW}$ are the  intrinsic dampings of the oscillators corresponding to the interlayer shear phonon and the CDW amplitude modes, respectively. $\eta=A_{CDW}/A_{ph}$ defines the ratio of the Raman amplitudes of the two Raman scattering channels. $\nu$ characterizes the coupling strength between the two modes. See SI Sec IV for further discussion of the Fano lineshape analysis.\\

\subsection*{Analysis of the pump-probe response}
We have implemented a convolution of fast rise, slow decay and overdamped oscillator to fit the transient time-resolved reflectivity as follows.
\begin{align}
    \frac{\Delta R}{R_0}&=A(\mathrm{erf}(\frac{t-t_0}{t_{rise}})+1)\ast(\mathrm{exp}(-\frac{t}{\tau_{decay}})+B)\\
    &\ast (\mathrm{exp}(-\gamma t)cos(2\pi f t+\phi)+C)
\end{align}
$t_0$ and $t_{rise}$ are respectively the onset and rise time of the ultrafast response.\\
$\tau_{decay}$ represents the relaxation time of the decay dynamics. $B$ is an offset to the decay, which physically represents a heating term. \\
$f$ is the frequency of the overdamped oscillation, and $C$ is an offset to the damping.
The intrinsic frequency of oscillation, $f_0$ is $f=\sqrt{f_0^2+\gamma^2}$, where $\gamma$ is the damping factor.\\
\section*{Author Contributions}
S.K., M.E. and A.K.N.M. contributed equally to this work. G.H., S.Ka. and N.P. conceived and designed the experiments. S.K. and A.K.N.M performed the ultrafast measurements, and analyzed the data. L.F. has performed the THz-THG measurements. S.K. and M.E. performed the Raman measurements and the subsequent analysis. T.C and V.G measured the electrical transport of exfoliated flakes. F.L.S and R.A supported the Raman measurements. S.K., M.E., and T.B discussed the acquisition and analysis of the Raman spectra. 
 The crystals were provided by S.C. and C.F.. M. A-H performed the specific heat measurements that provided additional support to the data. D. M., F.T., M. A-H, and K. N. validated and discussed the results with G.H., S.Ka. and N.P. S. K.,G.H.,S.Ka., and N.P. wrote the manuscript. All authors discussed the manuscript.
\section*{Acknowledgements}
N.P. acknowledges the partial funding by the European Union (ERC-CoG,
3DCuT, 101124606), by the Funded by the Deutsche Forschungsgemeinschaft (DFG, German Research Foundation): DFG 460444718   , DFG 512734967,    DFG   452128813,  DFG  539383397.
G.H. acknowledges financial support through start-up funding from IFW Dresden.
We acknowledge support from the Deutsche Forschungsgemeinschaft (DFG) through the Würzburg-Dresden Cluster of Excellence on Complexity, Topology and Dynamics in Quantum Matter—ctd.qmat (EXC 2147, Project No. 390858490) and Deutsche Forschungsgemeinschaft (DFG), SFB 1143 (project id 247310070).
S.Ka. acknowledges funding by the European Union (ERC, T-Higgs, GA 101044657). (Views and opinions expressed are however, those of the author(s) only and do not necessarily reflect those of the European Union or the European Research Council Executive Agency. Neither the European Union nor the granting authority can be held responsible for them). F.T. and D. M. would like to acknowledge the PNRR MUR
Project PE0000023 NQSTI. The authors thank Dr. Amit Pawabke for valuable discussions and insights. 
G.H. gratefully acknowledges the support from Dr. Matthias Finger of Oxford Instruments (WITec).
We are thankful to Prof. Nabeel Aslam and Martin Marcel from the University of Leipzig, Germany, for discussions.
\bibliography{bibliography}
\renewcommand{\thefigure}{S\arabic{figure}}
\clearpage
\onecolumngrid

\begin{center}
{\large\bfseries Supporting Information: Dynamics of CDW \textit{puddles} in 2\textit{H}-NbSe$_2$}\\[0.4cm]
\end{center}

\vspace{0.5cm}
\setcounter{figure}{0}

\newpage
\section{{Electrical transport measurements}}
\makeatletter
\renewcommand{\fnum@figure}{\textcolor{black}{\figurename~\thefigure}}
\makeatother
\begin{figure*}[ht]
    \centering
    \makeatletter
\renewcommand{\fnum@figure}{\textcolor{black}{\figurename~\thefigure}}
\makeatother
    \includegraphics[width=0.5\linewidth]{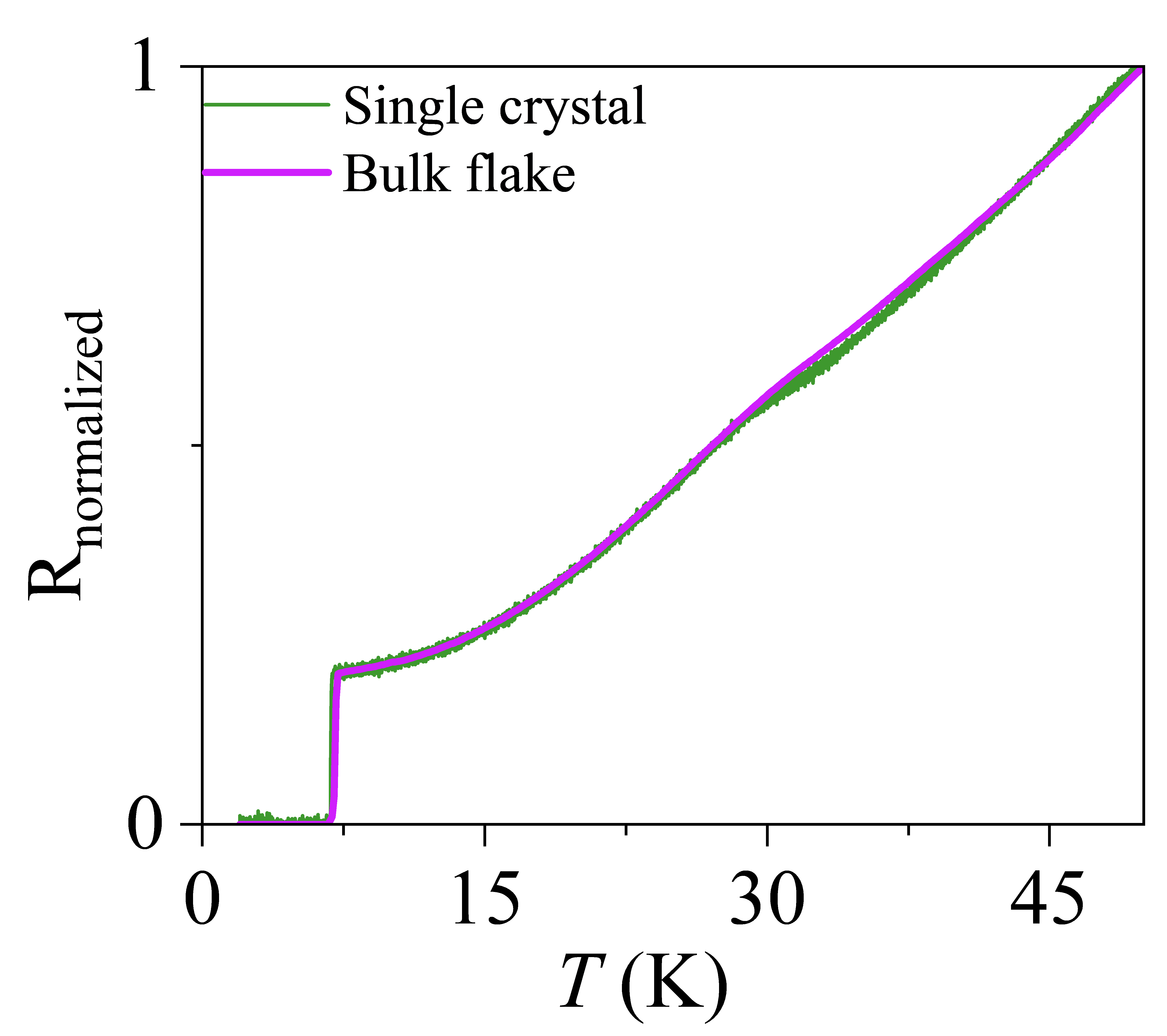}
    \caption{Four-probe resistance measurements in single crystals and exfoliated flakes of 2H-NbSe$_2$ on Si/SiO$_2$. The superconducting and CDW transition can be identified respectively at $\sim$6.8 K and 30 K}
    \label{RT}
\end{figure*}
\newpage
\section{Temperature-dependent ultrafast response}
\begin{figure*}[ht]
    \makeatletter
\renewcommand{\fnum@figure}{\textcolor{black}{\figurename~\thefigure}}
\makeatother
    \includegraphics[width=0.5\textwidth]{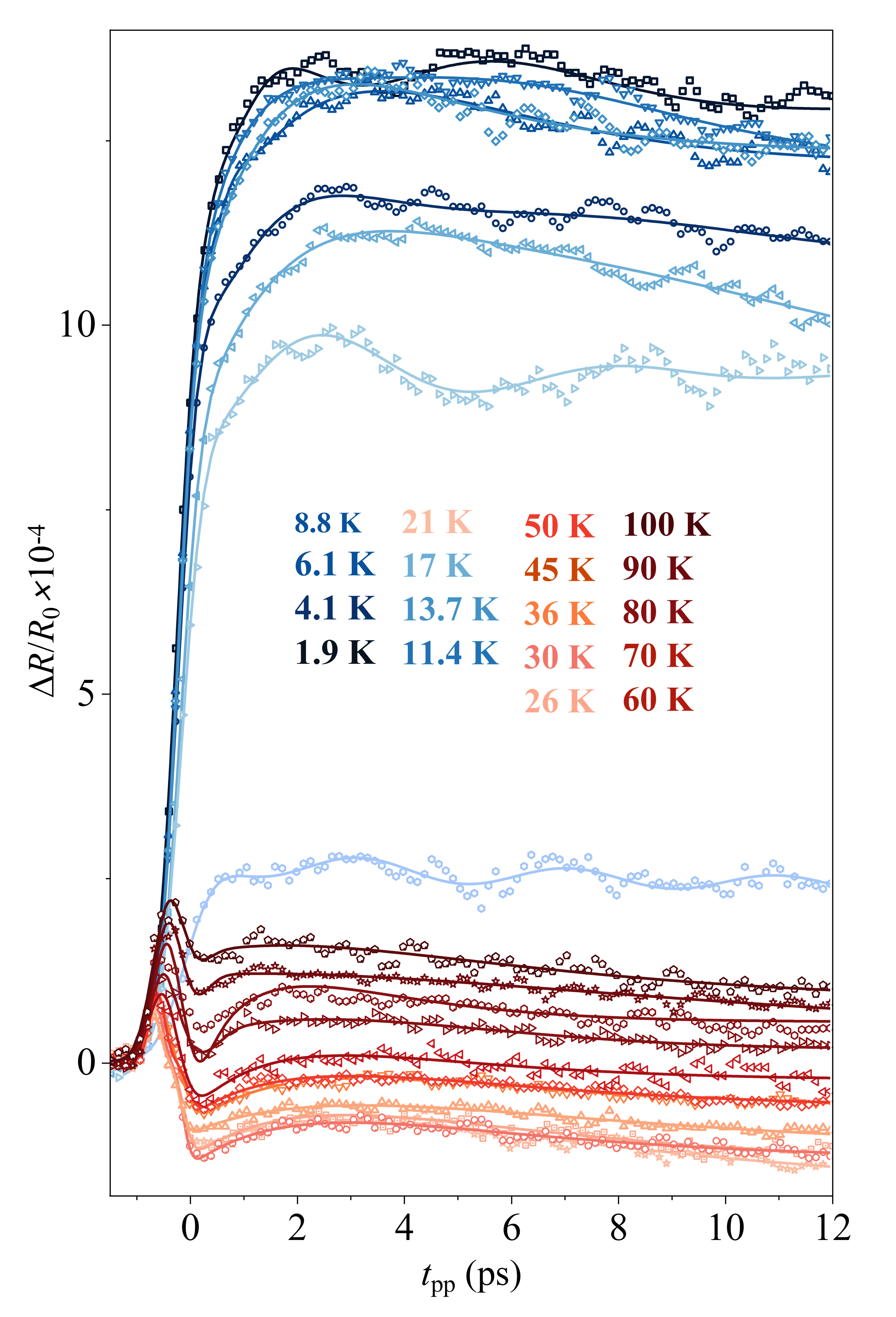}
    \caption{Time-resolved reflectivity change ($\Delta R$)
normalized with the reflectivity, R0, at zero pump-probe time-delay ($t_{pp}$) has been measured at
temperatures varying from 1.9 K to 100 K. The open points represent experimental data, and the
solid lines show the corresponding fits using the Eq. [2] of the main text.}
\label{ultrafast_T}
\end{figure*}
\section{Analysis of pump-probe measurements}
The Fano-coupled ph-CDW order that we present invokes an intertwined dynamics between lattice and charge degrees of freedom. In this description, the transient reflectivity cannot be captured by additive contributions from independent decay channels, as is commonly done in conventional pump-probe studies. The time-resolved reflectivity is instead described by a convolution of different decay dynamics:
a fast rise, slow decay, and an overdamped oscillator to fit the transient time-resolved reflectivity as follows.
\begin{align}
    \frac{\Delta R}{R_0}&=A(\mathrm{erf}(\frac{t-t_0}{t_{rise}})+1)\ast(\mathrm{exp}(-\frac{t}{\tau_{decay}})+B)\ast (\mathrm{exp}(-\gamma t)cos(2\pi f t+\phi)+C)
    \label{pp_analysis}
\end{align}
\textcolor{black}{$t_0$ and $t_{rise}$ are respectively the onset and rise time of the ultrafast response. We evaluated the intrinsic onset timescale of the dynamics, $t_{onset}$ by accounting for the finite temporal resolution of the experiment. With a cross-correlation of approximately 300 fs, we corrected the measured rise parameter by deconvolving the estimated IRF, yielding an IRF-corrected estimate of the onset time, $t_{onset}$.} \\
$\tau_{decay}$ represents the relaxation time of the decay dynamics. \\
\textcolor{black}{ A constant offset is required to define the electronic baseline associated with each channel. $B$ is an offset to the electronic decay, which physically represents a heating term. $C$ can be interpreted as an incoherent offset to the collective oscillation, without which the entire response would oscillate around zero. As coherence is lost, the oscillatory term decays to zero, whereas the incoherent contribution may remain finite.}

$f$ is the frequency of the overdamped oscillation, and $C$ is an offset to the damping.
The intrinsic frequency of oscillation, $f_0$ is $f=\sqrt{f_0^2+\gamma^2}$, where $\gamma$ is the damping factor.\\
\begin{figure*}[h] 
    \centering
    \makeatletter
\renewcommand{\fnum@figure}{\textcolor{black}{\figurename~\thefigure}}
\makeatother
    \includegraphics[width=0.6\textwidth]{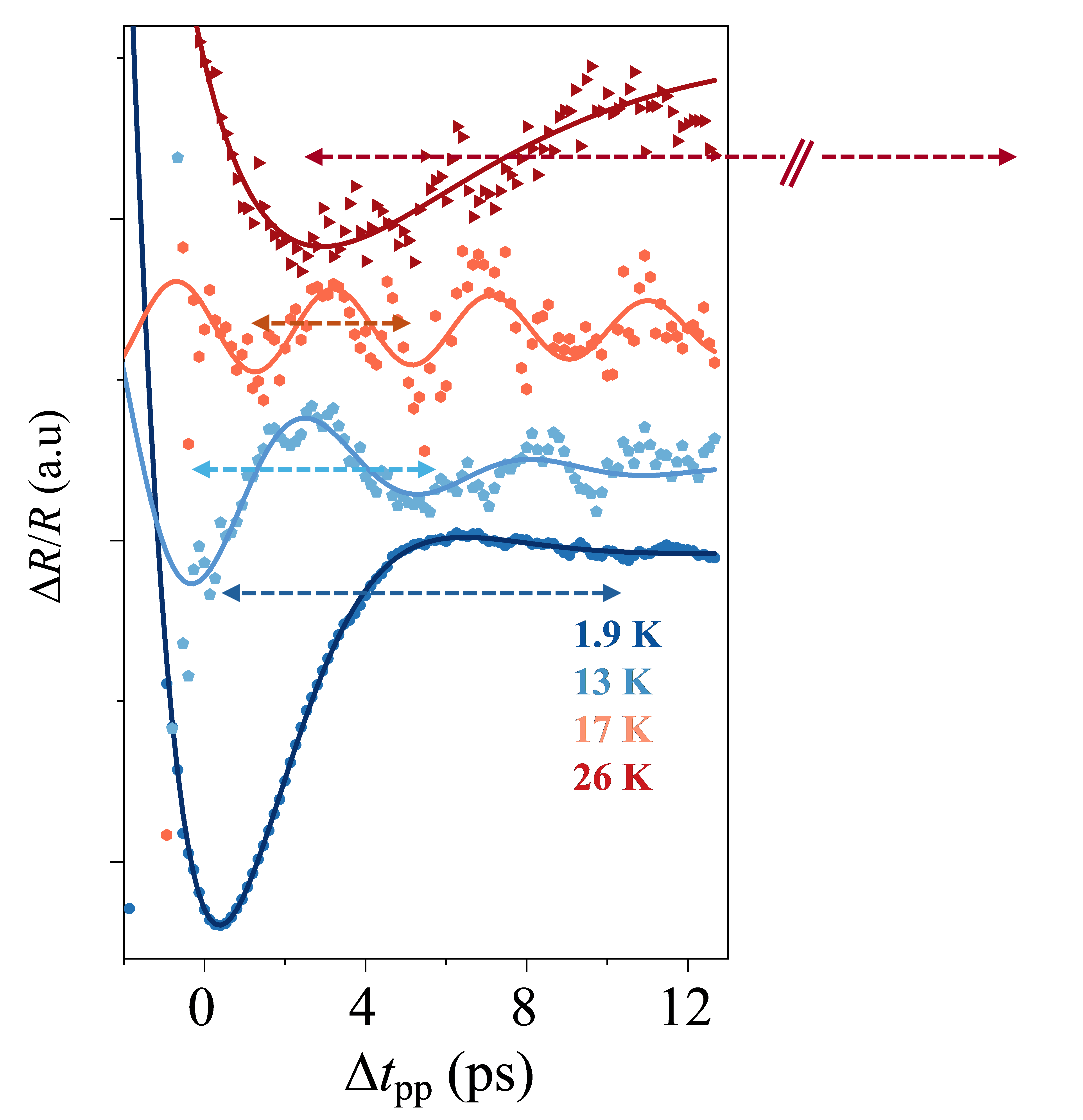} 
    \caption{\textbf{Overdamped oscillation in time-resolved reflectivity.} The normalized time-resolved reflectivity change, $\Delta R/R_0$ at 1.9~K, 13.7~K and 26~K is shown after deconvoluting with the ultrafast response and the slow electronic decay. The double-sided arrows represent approximate time-periods of the overdamped oscillation.}
    \label{oscillation}
\end{figure*}
 Fig.~\ref{oscillation} plots $\Delta R/R_0$ after deconvoluting with the measured data with the fast rise and slow decay, using Eq.~[1] at a few representative temperatures. We observe an overdamped oscillation, captured by the solid lines, that has approximately a time-period (estimated from the minimum close to zero time delay to the dip-like feature imposed on the slow decay) of 10 ps at 1.9~K. Close to 14 K, we observe the oscillation to be faster having an approximate time period of 5 ps, again becoming slower ($\ge12$~ps) at 26 K. This matches overall the $T$-dependence of the overdamped oscillation, $f_0$ that we show in Fig.~2(e) of the main text as well as in Fig.~\ref{specific heat}(e).
\begin{figure*}[ht] 
    \centering
    \makeatletter
\renewcommand{\fnum@figure}{\textcolor{black}{\figurename~\thefigure}}
\makeatother
    \includegraphics[width=0.8\textwidth]{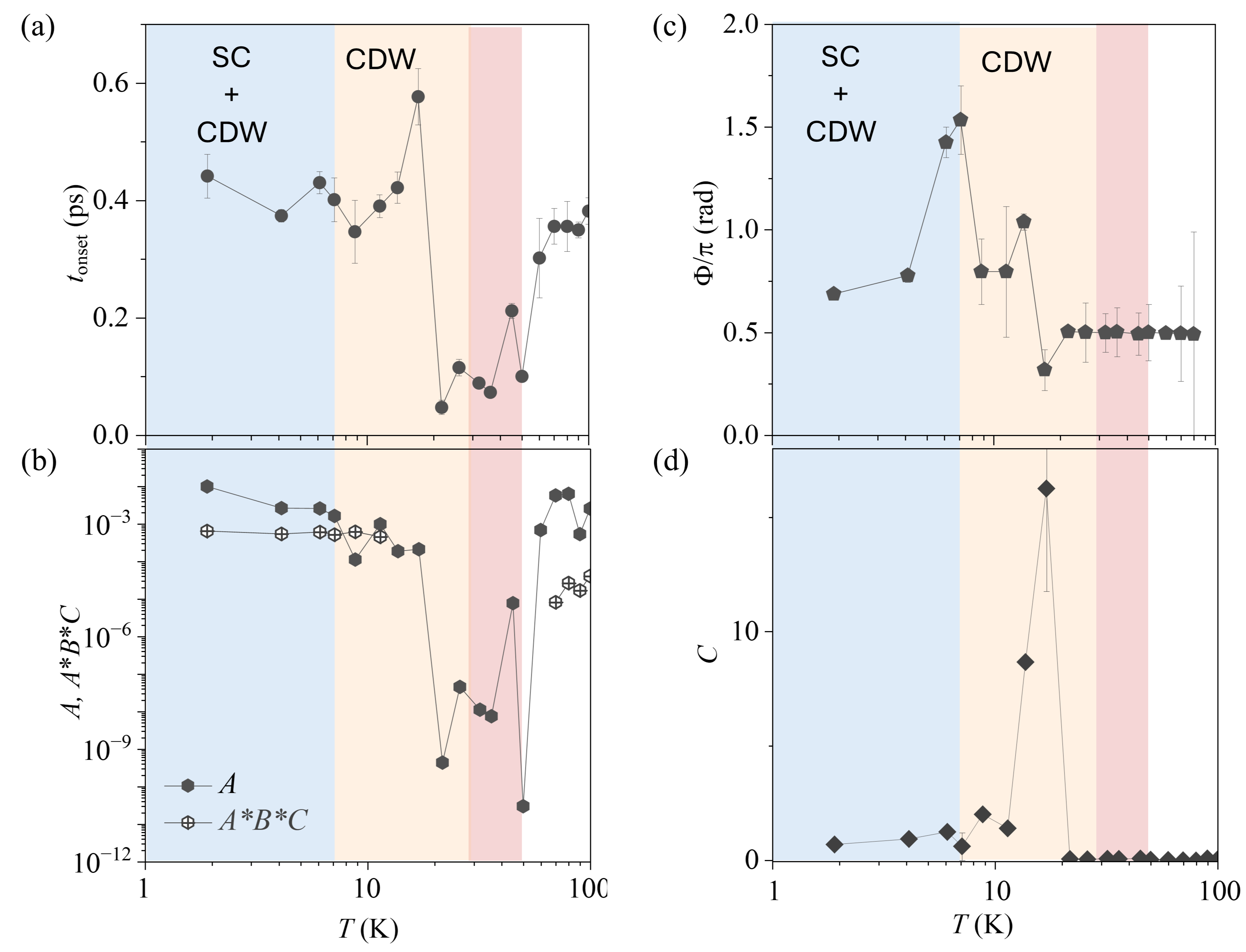} 
    \caption{\textbf{Parameters extracted from fits to the ultrafast response.} (a), (b), (c) and (d) plot the $T$-dependence of the onset timescale of the ultrafast response, $t_{rise}$, the amplitude ($A$) of the ultrafast response, as well as long-lived reflectivity ($A*B*C$), phase, $\Phi$ of the coherent oscillation, and the incoherent offset $C$, respectively at different temperatures. The blue, beige, and red regions are marked based on the transport characterization and Raman spectra.}
    \label{fit_parameters_PP}
\end{figure*}
Fig.~\ref{fit_parameters_PP}(a) plots the rise time of the ultrafast response, $t_{rise}$ at different temperatures. 
$t_{\mathrm{rise}}$ increases across $\sim 17$~K at lower temperatures, indicating a slowing down of the onset of the ultrafast response, which is quantitatively similar to ~\cite{PhysRevB.108.235160}, where this is attributed to CDW order-parameter suppression with higher temperature. 
However, a noticeable increase at $T\sim17$~K (and not at $T_{\mathrm{CDW}}$), as well as a drop at $\sim45$~K, cannot be entirely explained from this. We believe that below 50 K, an \textit{incoherent} puddle dynamics sets in, which shows a coherent oscillation only below $\sim17$~K. The order-parameter suppression is likely more effective in the {incoherent} puddle phase, whereas in the coherent regime, the collective oscillations slow down the onset. Interestingly, $t_{rise}\approx0.5$~ps, which is roughly $1/4$th of the CDW amplitude mode in NbSe2 at 35-40 cm$^{-1}$~\cite{hellmann2012time,doi:10.1126/science.1241591,PhysRevLett.83.800,
PhysRevB.99.235121}.  
Fig.~\ref{fit_parameters_PP}(b) shows the overall amplitude of the response (A) and the long-lived state (A*B*C), extracted from the fits in Fig.~\ref{ultrafast_T}, as a function of temperature. Both of them drop sharply close to 50~K and rise again at 17~K.
This is the same temperature regime of the incoherent puddles, which thus have the fastest rise and the smallest response.

\textcolor{black}{As shown in Fig. S3(c), $\Phi$ remains nearly constant above ($\sim17$)~K but undergoes a pronounced change in the same temperature range where we observe anomalies in the rise time, decay time, oscillation frequency, and Raman Fano parameter. This coincidence strongly suggests that the phase shift is linked to the emergence of the low-temperature collective state rather than to the excitation mechanism itself.
Within our interpretation, the observed oscillation originates from a hybrid phonon-CDW mode. In such a coupled system, the measured phase is determined not only by the excitation process but also by the relative contribution and coupling of the constituent degrees of freedom. For example, nonlinear THz studies on cuprates \cite{chu2020phase,chu2023fano,arpaia2023signature} show that the emergence of a Fano coupled order is accompanied by a characteristic phase shift. Similar measurements on 2H-NbSe2 \cite{PhysRevB.108.L100504} also indicate a phase shift from the Fano coupling of SC and CDW. While the microscopic origin may differ from the present case, these studies illustrate that the phase of the collective response is often a particularly sensitive signature of coupled orders. }

\textcolor{black}{The temperature dependence of $C$, shown in Fig. S3(d), reveals that it is finite only in the temperature range where the collective response is observed and rapidly approaches zero when the collective component disappears above ~17 K. Thus, the disappearance of C tracks the disappearance of the coherent response itself. This behaviour indicates that C is intrinsically linked to the collective phonon-CDW dynamics and should be viewed as an incoherent counterpart of the same coupled state.}

\textcolor{black}
{Below, we discuss the internal consistency of the intertwinned dynamics based on the fitting parameters.}
\subsection{Decay behaviour extracted from overdamped oscillatory model}

\textcolor{black}{
The extracted $\tau_{\mathrm{decay}}$ values shown in Fig.~2(d) in the main manuscript are about $30-100$ fs above 17~K, rise to $\geq$ 1~ns at 17 K, range in the order of $\sim 100$ ps until 11 K, where it drops down to $100$~fs. Below Tc, it again increases up to the order of $1$~ps. These values essentially span the hierarchy of relaxation scales, which has been established independently by Anikin \textit{et. al.} \cite{PhysRevB.102.205139} using broadband pump–probe spectroscopy of 2H-NbSe$_2$. (See Fig.~\ref{A1g_peak_comparison}(a).) Their global analysis involved a multi-exponential decay model, resolving characteristic times of approximately 100 fs, 1 ps, 100 ps and 1 ns. These are associated, respectively, with electron–electron thermalization, electron–phonon relaxation, anharmonic phonon–phonon relaxation and heat dissipation. }

\textcolor{black}{Above 17 K, where our extracted relaxation time ($\sim30-100$~fs) is extremely fast and IRF-limited, it is reasonable to associate the response primarily with a fast electronic relaxation process. Below 17 K, the increase to $100$~ps-$1$~ns, implies that the electronic perturbation becomes dynamically coupled to a much slower CDW/lattice relaxation pathway, so that the relaxation of the optically observed electronic response inherits the slow timescale of that coupled subsystem. Interestingly, the effective relaxation time does not simply increase continuously below 17 K. It first reaches few nanosecond regime close to 17 K and subsequently settles on hundreds of ps scale. These two scales are comparable to the slow relaxation scales independently identified by Anikin \textit{et. al.}~\cite{PhysRevB.102.205139}, namely thermal dissipation ($\sim1$~ns) and anharmonic phonon–phonon relaxation ($\sim100$~ps), respectively. We do not identify our extracted decay times directly with either microscopic channel. Instead, we interpret this evolution as evidence that the dominant bottleneck of the optically excited state changes abruptly near 17 K from a fast electronic response above this temperature to a much slower collective/lattice-controlled relaxation below it. Such a redistribution of relaxation weight is qualitatively consistent with the emergence of a spatially heterogeneous CDW state in which the electronic dynamics become increasingly coupled to the lattice and pinning degrees of freedom.}

\textcolor{black}{Interestingly, on further cooling below this slow-relaxation regime, $\tau_{decay}$ decreases again to 100~fs before rising to orders of $1$~ps near $T_c$. This timescale is comparable to the electron–phonon relaxation time identified by Anikin et al~\cite{PhysRevB.102.205139}. This suggests that the slow collective/lattice bottleneck dominating the response between approximately 10 and 17 K loses relative dynamical weight upon further cooling, allowing the conventional electron–phonon relaxation of the photoexcited electronic subsystem to become dominant again below Tc. Such non-monotonic behaviour is qualitatively consistent with a regime of slowly fluctuating CDW configurations that becomes increasingly pinned or quasi-static at lower temperature, close to Tc due to the possible impact of SC on the (dominant) CDW order and commensuration~\cite{doi:10.1021/acs.nanolett.8b00237,PhysRevB.108.L100504}. This is consistent with a concomitant decrease of the Fano coupling parameter. We also point to the fact that Anikin et.al~\cite{PhysRevB.102.205139} also report enhancement in the anharmonic phonon-phonon relaxation channel close to 15 K, similar to our $T$-scale of the onset of the coupled dynamics.}

\textcolor{black}{This comparison (shown in Fig.~\ref{tau_decay_comparison}(a)) shows that the magnitude of the characteristic times obtained from our fits is compatible with relaxation processes already known to occur in 2H-NbSe$_2$. Importantly, a multichannel decay model does not capture the glassy dynamics (see also Sec I.B) because of the intertwined nature of orders. This shows the decay dynamics is governed by the dominant relaxation bottleneck that can change to electronic or electron-phonon or phonon-phonon or so on. Thus, the strong temperature evolution is not an enormous change of one elementary scattering time, but a redistribution of the dominant relaxation pathway within the coupled electronic–lattice–CDW system.}\\
\textcolor{black}{\textbf{Oscillatory component} \\
All of this intertwined dynamics leads to a slow overdamped oscillatory response at the onset of the glassy puddle dynamics, as we discuss in the main text.
This observation is also supported qualitatively by the recent THz-STM results of Sheng et al. \cite{sheng2024terahertz}, which directly demonstrate spatially heterogeneous, defect-pinned CDW phase dynamics in 2H-NbSe2, including a low-energy response near 0.15–0.16 THz. While these measurements do not establish a transition at 17 K, they provide an independent microscopic basis for associating a low-frequency, strongly damped response with heterogeneous CDW dynamics rather than with a set of independent electronic relaxation channels.
}
\begin{figure}[ht]
\makeatletter
\renewcommand{\fnum@figure}{\textcolor{black}{\figurename~\thefigure}}
\makeatother
    \centering
    \includegraphics[width=1\linewidth]{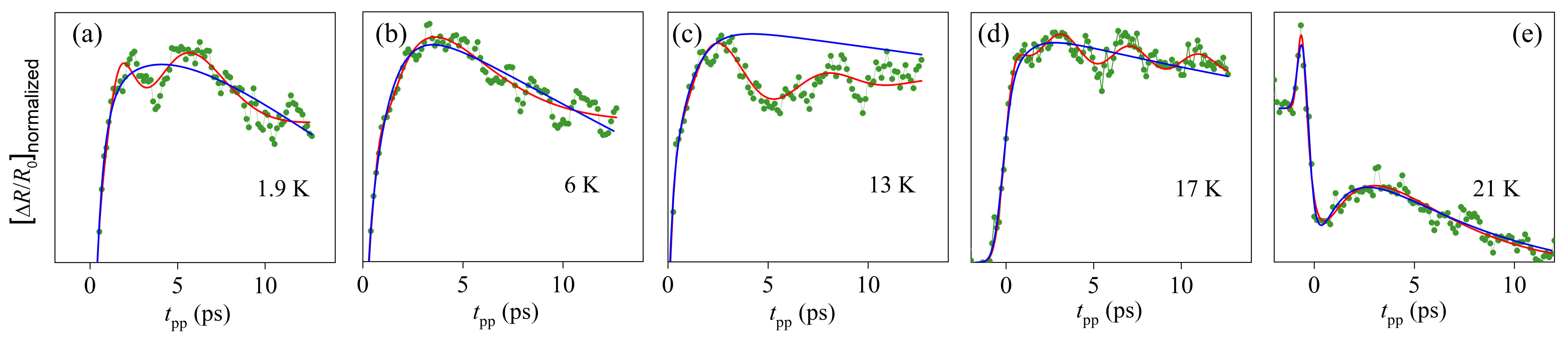}
    \caption{\textbf{Comparative analysis of the different fitting schemes of the measured transient reflectivity:} (a)-(e) show fits to the measured transient reflectivity with the overdamped oscillatory model (Eq.~[1]), and a multi-exponential decay (Eq.~[2]), represented by the red and blue solid lines, respectively, at different $T$. The plots have been zoomed appropriately to emphasize the difference between the two fitting schemes.}
    \label{Fits}
\end{figure}
\subsection{Decay behaviour extracted from multi-exponential model}
\textcolor{black}{We show below a comparison with the conventional multi-exponential decay analysis in the entire $T$-range. We have fitted the data using the following expression.
\begin{equation}
    \frac{\Delta R}{R_0}=
\mathrm{IRF} \otimes [A_1*exp(-(t-t_0)/t_1)+A_2*exp(-(t-t_0)/t_1)+A_3*exp(-(t-t_0)/t_3)+B].
\end{equation}
Based on the fits (some representative $T$s are shown in Fig.~\ref{Fits}), we note the following.\\
1.	The model naturally fails to capture the oscillatory component.\\ 
2.	Overlooking the oscillatory compoenent, we require three relaxation channels to capture the average behavior of the transient reflectivity. From the $T$-dependence of the fitted parameters (See Fig.~\ref{tau_decay_comparison}), we observe that several of the resulting relaxation constants become comparable in magnitude, which is unphysical unphysical based on earlier report by Anikin \textit{et.al.}~\cite{PhysRevB.102.205139}. This leads to substantial parameter correlation and large uncertainties. The shortest component repeatedly approaches or falls below the experimental temporal resolution and is consequently not quantitatively constrained. On the other hand, relaxation times extracted from the overdamped oscillatory model show cleaner $T$-dependence, and clear separation among the different orders of magnitude allowing the identification of different decay channels in comparison with Ref\cite{PhysRevB.102.205139}.\\ 
3.	More importantly, around 17 K the estimated relaxation times in all the channels show large simultaneous changes by orders of magnitude. None of the earlier reports shows such drastic enhancements of values in one channel. Hence, the enhancements that we observe here, in both the models, emerge from crossover in the dominating relaxation channel rather than an enhancement in the same channel. This mechanism cannot be captured by a multi-exponential decay model, whereas an overdamped oscillatory dynamics indeed shows this consistently. The sum of exponentials may apparently provide a representation of the averaged transient, especially when only a fraction of an oscillation is apparent in the time domain. But it does not signify physically true independent relaxation channels, and hence the estimation of the extracted parameters is unreliable in this case.\\
To summarize, the exponential decomposition of the transient reflectivity is either failing to capture the measured dynamics or has degeneracy in the relaxation channels, while the overdamped oscillatory fitting framework successfully captures the intertwined glassy dynamics and has (i) independent support from the known 2H-NbSe$_2$ relaxation hierarchy in terms of the extracted relaxation times, and (ii) the directly observed low-frequency ($\sim0.15$~THz) CDW dynamics. 
}

\begin{figure}
\makeatletter
\renewcommand{\fnum@figure}{\textcolor{black}{\figurename~\thefigure}}
\makeatother
    \centering
    \includegraphics[width=1\linewidth]{Fig_S6_n.png}
    \caption{\textbf{Comparison of relaxation times extracted by fitting the transient reflectivity with overdamped oscillatory component and multi-exponential decay.} (a) $\tau_{decay}$ is reproduced from Fig. 2c of the main manuscript. Four different scales of the time constants have been identified, as denoted by $\tau_1\sim100$~fs,$\tau_2\sim1$~ps,$\tau_3\sim100$~ps,$\tau_4\sim1$~ns . (b)-(d) show the multiple decay channels estimated by fitting the reflectivity with multi-exponential decay, as shown by the blue solid lines in Fig.~\ref{Fits}.}
    \label{tau_decay_comparison}
\end{figure}




\newpage
\section{Temperature-dependent Raman response}

\begin{figure}[ht]
    \centering
    \includegraphics[width=0.7\linewidth]{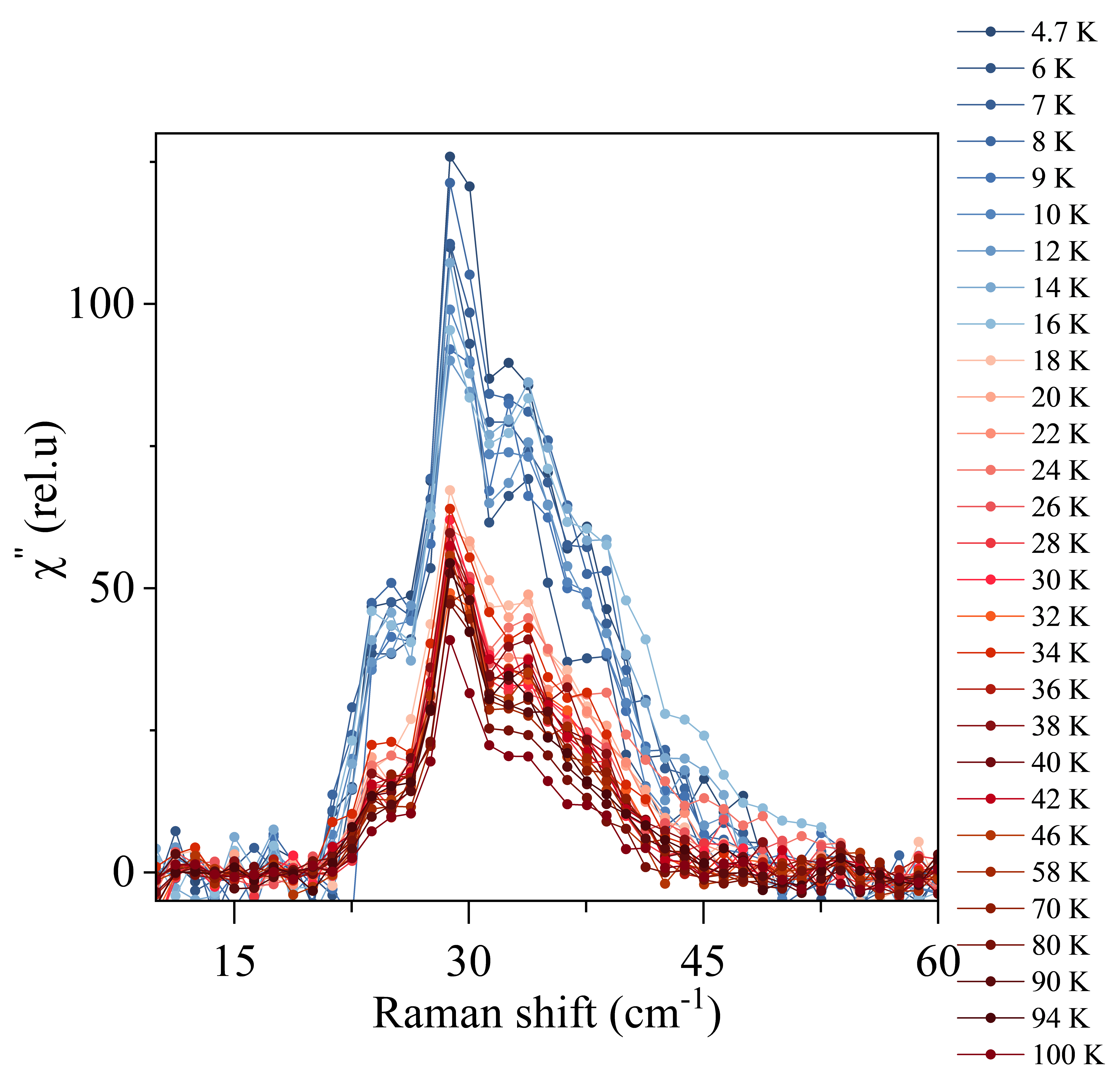}
    \caption{\textbf{Temperature-dependent Raman susceptibility}. Raman susceptibility ($\chi''$) measured on bulk-like (thickness$\sim60$~nm) exfoliated flakes of NbSe$_2$ at temperatures varying from 4.7~K to 100~ K in an unpolarized backscattering configuration is shown in the low-frequency spectral range up to 100 cm$^{-1}$.}
    \label{raman_T}
\end{figure}
Fig.~\ref{raman_T} shows the low-frequency Raman response for different temperatures.
We estimated the laser-induced heating effects to be less than 1 K.

\begin{figure*}
    \makeatletter
\renewcommand{\fnum@figure}{\textcolor{black}{\figurename~\thefigure}}
\makeatother
\includegraphics[width=0.45\textwidth]{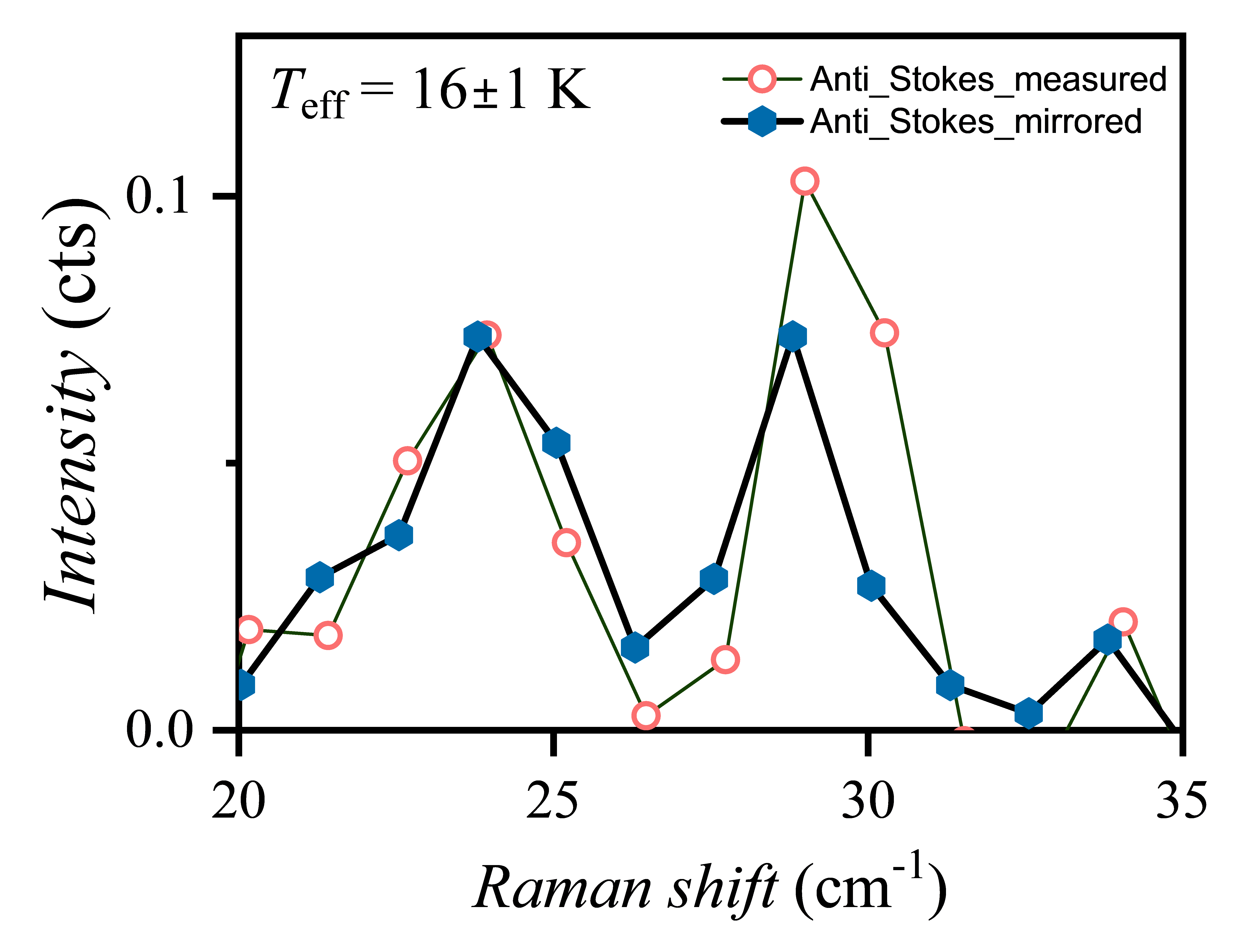}
    \caption{Measured Anti-Stokes spectrum and Anti-Stokes spectrum simulated from the Stokes intensity using Eq.~[2] are plotted.}
        \label{Stokes}
\end{figure*}

\textcolor{black}{Laser-induced heating is most commonly assessed through the effective temperature extracted from the Stokes/Anti-Stokes intensity ratio. However, in the present case, the low-energy response originates from a Fano-coupled phonon-CDW mode, for which the relation between the measured intensity and the thermal population is not well-understood in the literature. Consequently, applying a conventional Stokes/Anti-Stokes analysis directly to the hybrid mode may not yield a reliable estimate of the lattice temperature. To circumvent this issue, we instead consider the split mode near 25 cm$^{-1}$, which is well described by a Lorentzian lineshape in our Raman analysis [See discussion of Fig. 3 in manuscript] and therefore exhibits predominantly phononic character. Using the intensity ratio of the Stokes and Anti-Stokes components (following Eq.~2) of this mode after excluding the ph-CDW response, we estimate the effective temperature of the sample.}
\begin{equation}
\label{Tcalc}
    \frac{I_{As}(\Delta E)}{I_{As}(\Delta E)}=\frac{(E+\Delta E)^4}{(E-\Delta E)^4}\mathrm{exp}(-\frac{\hbar\Delta E}{k_B T})
\end{equation}
\textcolor{black}{We show in Fig.~\ref{Stokes} the measured anti-Stokes intensity and the estimated anti-Stokes from the Stokes data at 16 K using the expression below~\cite{bock1995laser, glier2025non}.}
\textcolor{black}{
We focus on this temperature because it represents the most critical temperature scale identified in the manuscript, where we claim the onset of the coupled collective dynamics. In addition, near this temperature range, the split mode around $\sim25$~cm$^{-1}$ is most clearly resolved, allowing the most reliable determination of the effective temperature from the Stokes/Anti-Stokes intensity ratio. We find that the measured Anti-Stokes response is reproduced by the mirrored Stokes spectrum using an effective temperature of $T_{\mathrm{eff}}\pm1$~K. Thus, within experimental uncertainty, the effective temperature agrees with the bath temperature.}
\section{Fano Analysis}
The expression we used to fit the Fano line shape in Raman spectra is derived from a Fano-type resonance that describes the coupling between two discrete
excitations with finite dampings~\cite{Klein1983, mialitsin2011fano, eklund1979analysis, blumberg1994investigation, Joe2006}. 
\begin{equation}
\begin{aligned}
f(\omega) = & \frac{A_{ph}^2\pi r(\omega) \left[ \left (\omega -  \omega_{CDW} + \nu \eta \right)^2 + \gamma_{CDW} \left( \nu^2 \pi r(\omega) + \gamma_{CDW} \right) \right]}{(\omega- \omega_{CDW}  - \nu^2 R(\omega))^2 + (\nu^2 \pi r(\omega) + \gamma_{CDW})^2} \\
& + \frac{A_{ph}^2\gamma^2_{CDW} \left( \nu R(\omega) + \eta \right)^2}{(\omega - \omega_{CDW} - \nu^2 R(\omega))^2 + (\nu^2 \pi r(\omega) + \gamma_{CDW})^2}
\label{fano_expression_final}
\end{aligned}
\end{equation}
where:
\[r(\omega)=\frac{\gamma_{ph}^2}{(\omega - \omega_{ph})^2 + \gamma_{ph}^2}\]
 and $R(\omega)$ is derived as Hilbert transform of $r(\omega)$.
Here, $\omega$ is the Raman-frequency shift. $\omega_{ph},\omega_{CDW}$ are the centre frequencies, $\gamma_{ph},\gamma_{CDW}$ are the  intrinsic dampings of the oscillators corresponding to the interlayer shear phonon and the CDW amplitude modes, respectively. $\eta=A_{CDW}/A_{ph}$ defines the relative oscillator strength of these modes.
Physically, we interpret this as the thermally-broadened ($\gamma_{ph}$) interlayer shear vibrational phonons to couple with the CDW amplitude mode and produce a damping to its sharp resonance, $\nu^2\pi r(\omega)$, from the Fano coupling in addition to an intrinsic broadening $\gamma_{CDW}$. 

\begin{figure}[ht]
    \centering
   \makeatletter
\renewcommand{\fnum@figure}{\textcolor{black}{\figurename~\thefigure}}
\makeatother
    \includegraphics[width=0.5\textwidth]{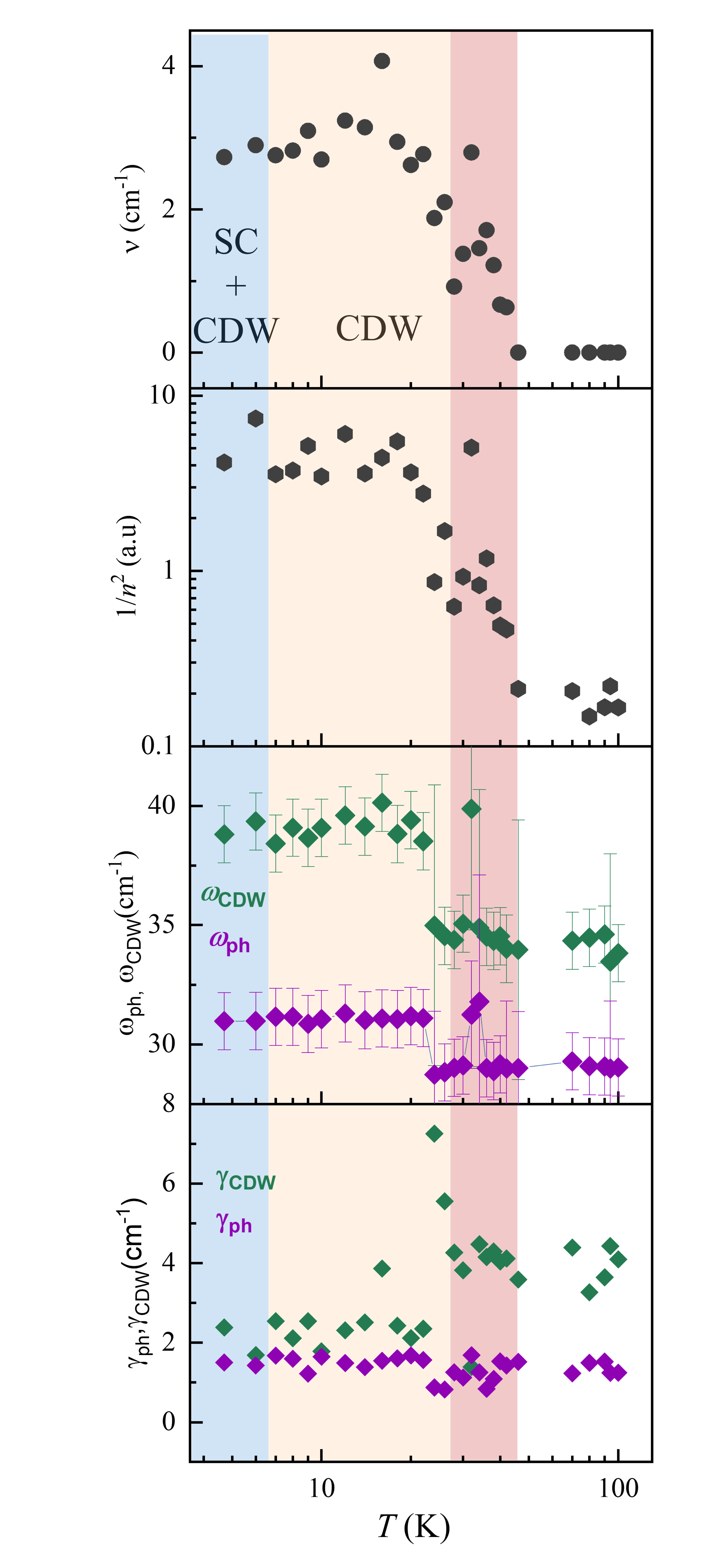} 
    \caption{\textbf{Parameters extracted by fitting the Raman response with the Fano linsehape.} (a), (b), (c) and (d) respectively plot the Fano coupling ($\nu$), relative oscillator strength between the phonon and CDW modes ($\eta$), centre frequencies ($\omega_{ph},\omega_{CDW}$) and damping ($\gamma_{ph},\gamma_{CDW}$).}
    \label{Fano_fit_parameters}
\end{figure}

Fig.~\ref{Fano_fit_parameters} plots the $T$-dependence of the parameters extracted from the Fano fits to the Raman response with (a), (b), (c) and (d) respectively plotting $\nu$, 
$\eta$, ($\omega_{ph},\omega_{CDW}$), and ($\gamma_{ph},\gamma_{CDW}$).
All these parameters show noticeable changes at $\sim17$~K, which we note is the onset of the coherent CDW-puddle dynamics. The $T$-dependence of $\nu$ in Fig.~\ref{Fano_fit_parameters}(a), discussed in the main text, captures an (i) early onset of puddles, which we define as an incoherent puddle regime below 50~K, (ii) a maximum strength close to $17$~K, that marks the onset of a {coherent}-puddle phase emerging likely with the competition between different orders of CDW commensuration in the material. $1/\eta^2$ (Fig.~\ref{Fano_fit_parameters}(b)) shows a saturation below $T\sim17$~K, which is in line with a stable coupling between the lattice and the CDW order, leading to the formation of this coherent phase. Both $\omega_{ph}$ and $\omega_{CDW}$ in Fig.~\ref{Fano_fit_parameters}(c) increase below 17 K, accompanied by a decrease in $\gamma_{CDW}$ (Fig.~\ref{Fano_fit_parameters}(d)), consistent with a transition in the system. We conclude based on these that the transition at 17 K marks the evolution of CDW puddles from fluctuating, weakly coupled domains into a more rigid, phase-locked configuration.
This is also in line with the hardening of the interlayer shear vibration, as the puddles become pinned and more interlayer-correlated.
\textcolor{black}{We note that though the mode frequencies, $\omega_{ph}$ and $\omega_{CDW}$ and damping constants, $\gamma_{CDW}$ and $\gamma_{ph}$ show a sharp change close to 23 K, we do not interpret this as evidence for a distinct transition temperature. 
Instead, within the experimental resolution of 1.2~cm$^{-1}$, we view the temperature interval between approximately 17 K and 25 K as a crossover regime in which the coupled phonon-CDW response evolves continuously. The identification of 17 K as the onset of the coherent collective regime is based on the correlated anomalies observed across multiple independent observables and experimental probes rather than on any individual Raman fitting parameter alone.}\\
\textcolor{black}{Finally, we would like to emphasise that the low-energy Raman response in 2\textit{H}-NbSe$_2$ remains surprisingly unresolved despite decades of study. A notable aspect of the reported CDW amplitude mode is that it is unusually broad, with linewidths of the order of tens of cm$^{-1}$, which is itself unexpected for a well-defined collective excitation, for example, in comparison to the Higgs mode identified in the very same material. This broad response has been recognized since the early Raman studies of 2\textit{H}-NbSe$_2$, yet its microscopic origin remains debated. While the amplitudon was first identified in Ref. \cite{PhysRevLett.37.1407}, subsequent reports have shown significant variation in both the position and linewidth of the CDW-related feature. Some studies report a broad response centred near 45 cm$^{-1}$ with a linewidth approaching 20 cm$^{-1}$
\cite{PhysRevLett.37.1407,PhysRevB.89.060503}, whereas others identify a feature closer to 35 cm$^{-1}$ \cite{xi2015strongly, lin2020patterns,shi2026enhancedchargedensitywaveordersuppressed}. Likewise, the relative intensity between the interlayer shear phonon and the CDW-related response varies substantially across the literature.
These discrepancies indicate that the low-energy Raman response of 2\textit{H}-NbSe$_2$ remains incompletely understood, despite the extensive attention the material has received over the past several decades. Indeed, much of the recent attention has focused on the superconducting Higgs mode and its coupling to the CDW state, focusing primarily on temperatures upto~10 K, whereas a detailed temperature-dependent investigation of the CDW-related Raman response itself has remained largely unexplored.
The hybridization between the interlayer shear phonon and the CDW response, that we discuss, explains both the unusually broad lineshape and the apparent variation in the reported peak position between 35 and 45 cm$^{-1}$. In fact, the presence of a Fano-type interaction between the phonon and CDW response was already pointed out by Mialitsin et al.\cite{mialitsin2011fano}. What has been missing, however, was a systematic temperature-dependent investigation of this coupling and its relation to the emergence of a distinct collective dynamics. By combining temperature-dependent Raman spectroscopy with ultrafast measurements, the present work provides such a framework and demonstrates that the evolution of the Fano coupling parameter is directly linked to the onset of a new collective response. Within this framework, the observed Raman response is determined not only by the intrinsic CDW excitation but also by the relative spectral weight of the phononic and CDW components. Consequently, modest variations in the relative strengths of these coupled contributions can naturally produce Raman responses that appear centred closer to either 35 or 45 cm$^{-1}$, thereby providing a unified interpretation of the existing literature.
}
\clearpage
\newpage
\section{High-frequency Raman response in N\lowercase{b}S\lowercase{e}$_2$}
\begin{figure}
\makeatletter
\renewcommand{\fnum@figure}{\textcolor{black}{\figurename~\thefigure}}
\makeatother
    \centering
    \includegraphics[width=1\linewidth]{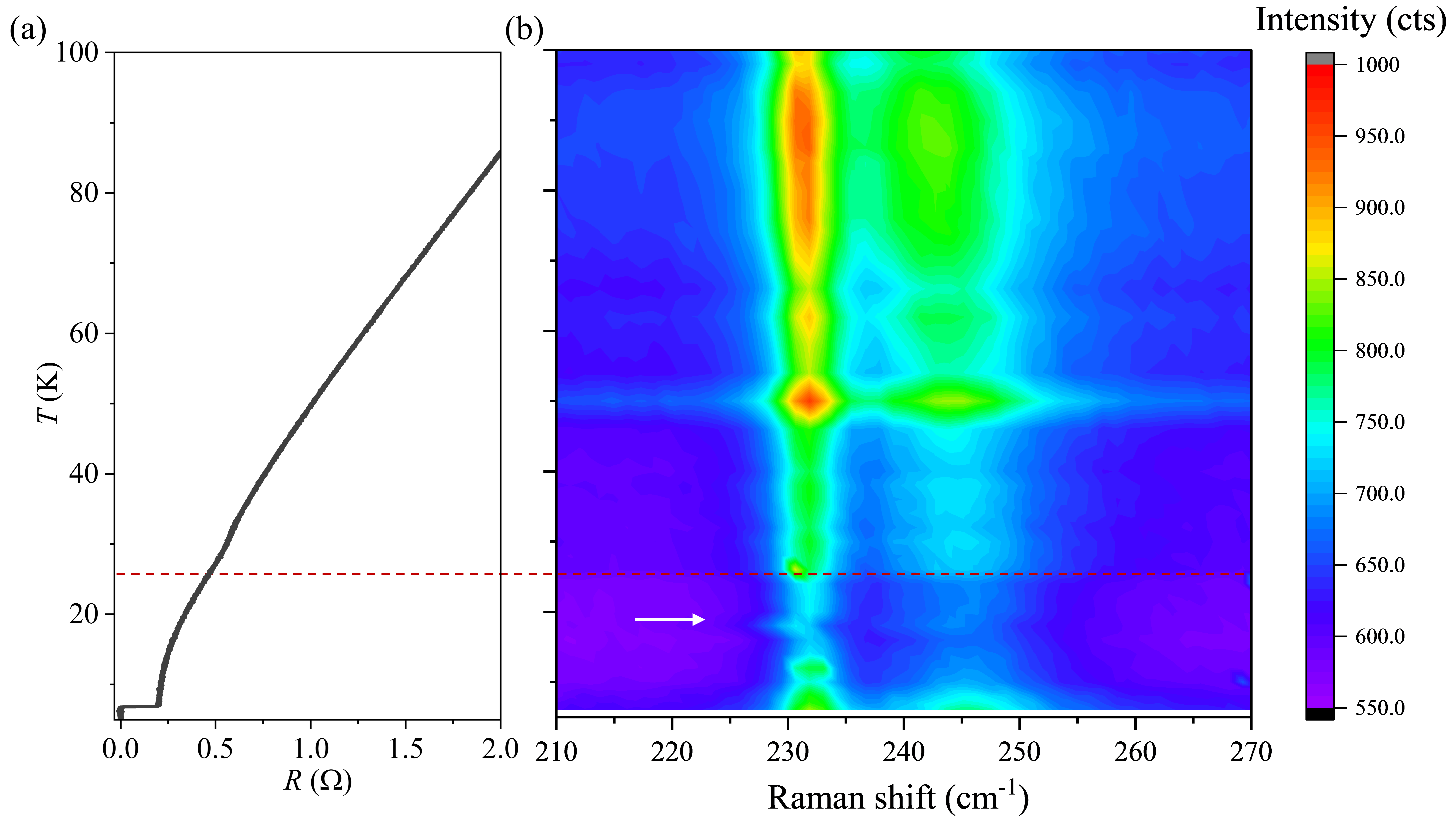}
    \caption{\textbf{High-frequency Raman Response}: (a) Electrical transport characteristics reproduced from Fig.~\ref{RT}. (b) $T$-dependence of the Raman Intensity between $210-270$~cm$^{-1}$. The red dotted line and white solid arrow, respectively, indicate the CDW transition at $T_{CDW}\approx28$ and the onset of the puddle dynamics at $17$~K. }
    \label{high_freq_color_plot_high_fluence}
\end{figure}

\textcolor{black}{
The high-frequency Raman response measured under the same fluence, shown in Fig.~\ref{high_freq_color_plot_high_fluence}, retains a distinct signature of the conventional CDW transition near $T_{\mathrm{CDW}}\approx 28$~K, while a separate characteristic change occurs near 17 K. 
The simultaneous observation of distinct features near 17 K and $T_{\mathrm{CDW}}$ also rules out the interpretation of the 17 K anomaly as a laser-heating induced shift of the ordinary CDW transition. \\
The frequency and linewidth of these high-frequency phonons also reflect the electronic environment and electron–phonon coupling, and their temperature evolution therefore provides an independent view of the coupled electronic–lattice response. The concomitant evolution of both high-frequency optical phonons in our measurements indicates that the lattice sector itself evolves across the CDW regime. The A1g mode is particularly useful quantitatively because its linewidth has previously been shown to be sensitive to CDW-related electron–phonon coupling in 2H-NbSe2 \cite{https://doi.org/10.1002/adma.202003746}. More generally, recent temperature-dependent Raman studies of layered CDW materials have shown that the evolution of multiple Raman-active phonons can track CDW phase boundaries and reflect changes in the coupled electronic and lattice degrees of freedom \cite{pxpf-6bsv}. Hence, the clear dependence of the A1g peak as well as E2g peak at 28 K and 17 K proves the absence of any significant laser-induced heating. Our data also show a broadening of the peaks starting again at 11 K, where we observe a drop in the $\tau_{decay}$, $f_0$, as well $\nu$. 
Hence, all the $T$ scales can be clearly identified in our data at the equilibrium values.
}

\begin{figure*}[ht]
    \includegraphics[width=0.75\textwidth]{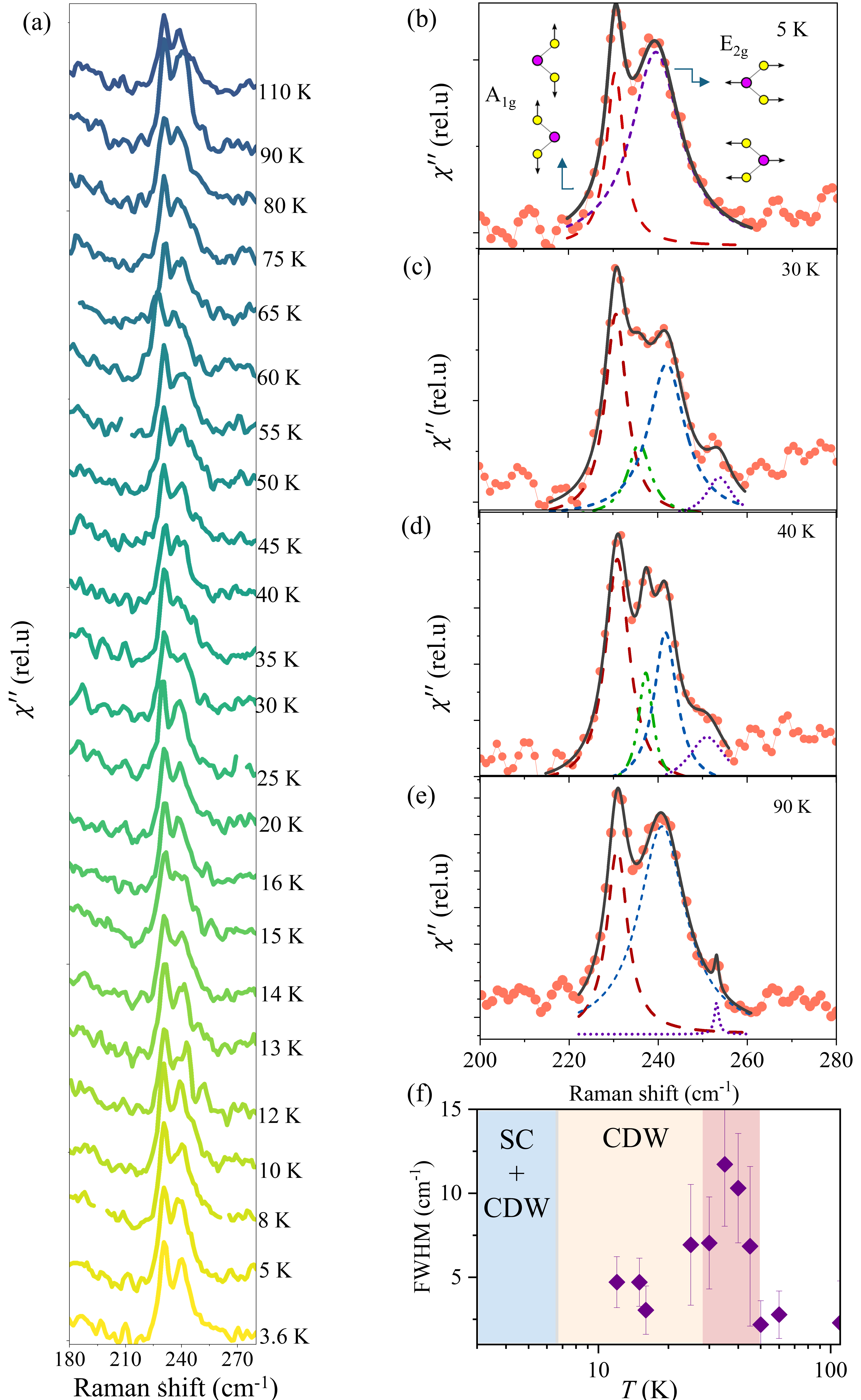} 
    \caption{\textbf{High-frequency Raman response in flakes of 2\textit{H}-NbSe$_2$.} (a) $T$-dependence of the Raman susceptibility, $\chi^{"}$ within the spectral range $180-270$~cm$^{-1}$ is shown. (b)-(e) plot $\chi^{"}$ at $T\sim$~5,~30,~40,~and~90~K respectively. The points represent the experimental data, and the solid lines show the corresponding fits, as discussed in the text. (f) shows the $T$-dependence of the the full-width half maximum (FWHM) of the shoulder peak (represented by the purple-dotted line).}
    \label{high_frequency Raman peaks}
\end{figure*}

\begin{figure}
\makeatletter
\renewcommand{\fnum@figure}{\textcolor{black}{\figurename~\thefigure}}
\makeatother
    \centering
    \includegraphics[width=0.5\linewidth]{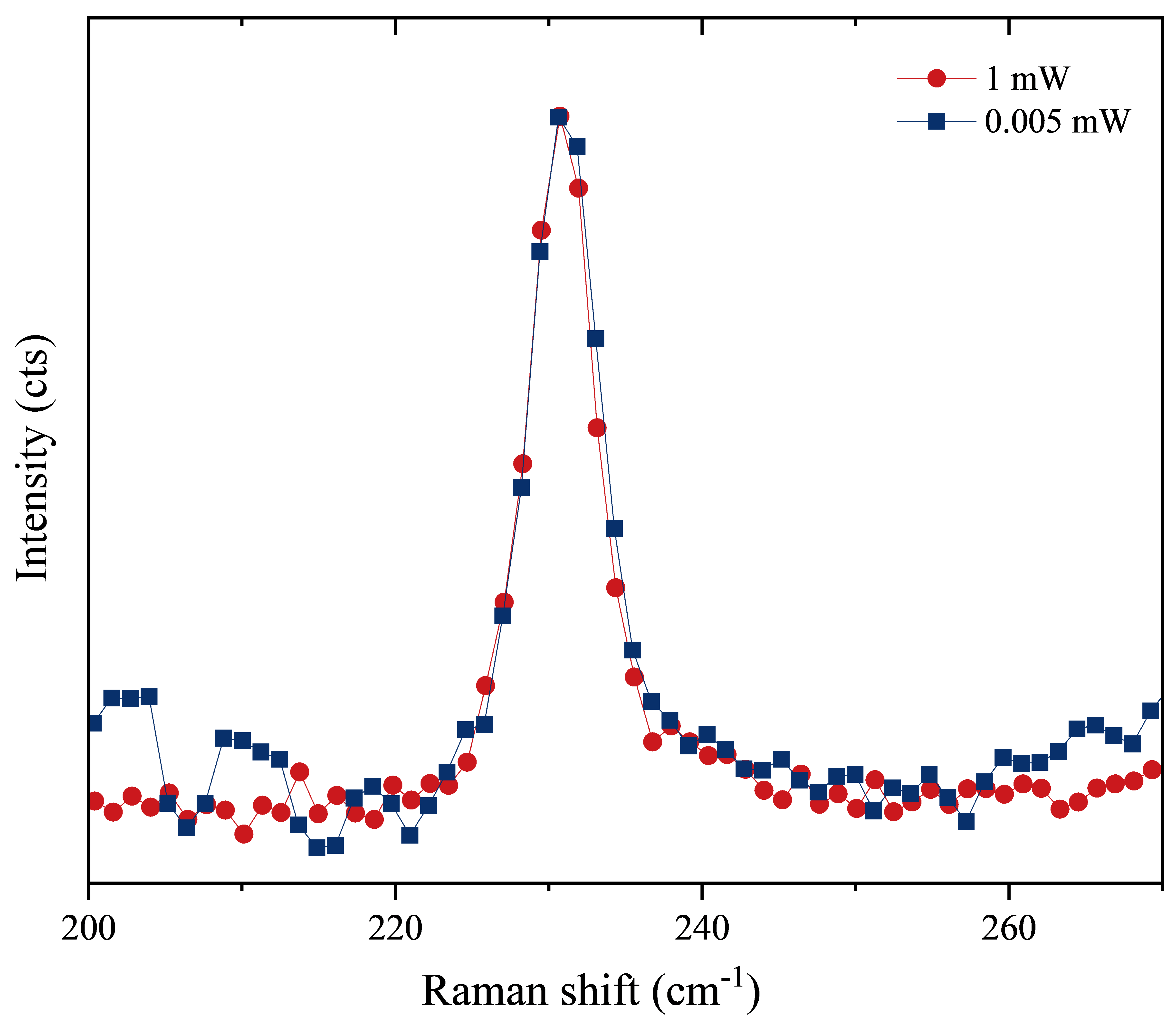}
    \caption{A$_{1g}$ peak in 2H-NbSe$_2$ at two fluences of 1 mW and 5 $\mu$W at 16 K.}
    \label{A1g_peak_comparison}
\end{figure}

Fig.~\ref{high_frequency Raman peaks}(a) shows the $T$-dependent Raman measurements on NbSe$_2$ within the spectral range $\sim180-270$~cm$^{-1}$ performed at low power of 5~$\mu$W. Two distinct Raman modes can be clearly identified, around 230 and 240 cm$^{-1}$, representing the well-known A$_{1g}$ and E$_{2g}$ modes, corresponding to the in-plane (IP) and out-of plane (OP) vibrations, respectively~\cite{PhysRevB.89.060503,xi2015strongly}. We emphasize here a shoulder peak close 250 cm$^{-1}$, which can be identified below 90~K. The measurements have been performed in an unpolarized configuration with a lower power of the laser, where we find the shoulder peak to be resolved better. Each spectrum has been integrated over 20 minutes and averaged over 5 accumulations to improve the signal-to-noise ratio. 

We plot the Raman spectra at three distinct $T$s separately in Fig.~\ref{high_frequency Raman peaks}(b)-(c) to better understand this. Fig.~\ref{high_frequency Raman peaks}(b) shows that at 5 K, the data show two clear peaks at 230 cm$^{-1}$ and 240 cm$^{-1}$, corresponding to the IP and OP vibrations respectively, whereas at higher $T$s in addition to the IP A$_{1g}$ peak 230 cm$^{-1}$, two peaks around 240 cm$^{-1}$ (specifically at $236$~cm$^{-1}$ and $242$~cm$^{-1}$) and another peak at $\sim250$ cm$^{-1}$ can be fitted. 
The peaks centered at $236$~cm$^{-1}$ and $242$~cm$^{-1}$ (represented by the blue and green dashed lines respectively) can probably result from a splitting of the doubly degenerate E$_{2g}$ peak (identified at $240$ cm$^{-1}$ at 5~K) predominantly close to the CDW transition.
However, the peak close to 250 cm$^{-1}$ remains elusive, though earlier reports \cite{PhysRevLett.37.1407} have also identified this. 
We plot the full-width half maximum (FWHM) of this peak in Fig.~\ref{high_frequency Raman peaks}(f), showing the broadening to be maximum between 30-60 K. 

Based on a clear change close to 50 K, we propose that this arises as short-range CDW puddles break translational symmetry and activate folded phonons incoherently. At lower temperatures, long-range coherence selects a single phase, and hence the peak is suppressed. This is also in line with the onset of an incoherent CDW puddles at this temperature discussed in Sec II-III and in the main text.

\clearpage
\textcolor{black}{We have explicitly examined whether the Raman excitation could induce an appreciable steady-state photodoping of the sample and thereby influence the observed Raman response or the characteristic temperature scales discussed in the manuscript. The high-frequency out-of-plane A$_{1g}$ phonon provides a useful experimental probe of changes in the electronic environment. In semiconducting TMDs, carrier doping is known to renormalize the A$_{1g}$ phonon frequency and linewidth through electron–phonon coupling~\cite{PhysRevB.85.161403}, and analogous Raman signatures have been used to identify photoinduced changes in carrier density \cite{miller2015photogating}. In 2H-NbSe$_2$, the A$_{1g}$ response is additionally coupled to the electronic/CDW environment. In particular, previous studies \cite{https://doi.org/10.1002/adma.202003746} have shown that interfacial charge transfer modifies the CDW state and its coupling to the A$_{1g}$ phonon, with the A$_{1g}$ linewidth providing a sensitive measure of the resulting CDW modification. 
To test for such laser-induced changes in our experiment, we compare the A$_{1g}$ response measured at 1 mW with that measured at 5 $\mu$W. Despite this large variation in photon flux, within our experimental resolution we observe no significant shift of the A$_{1g}$ phonon frequency and no systematic modification of its linewidth or lineshape. We therefore find that laser-induced photodoping is negligible on the scale relevant to the Raman features and temperature-dependent collective dynamics discussed in this work.
}
\newpage
\section{Additional specific heat and THz-THG measurements on 2\textit{H}-N\lowercase{b}S\lowercase{e}$_2$ crystals}
\begin{figure*}[ht]
   \makeatletter
\renewcommand{\fnum@figure}{\textcolor{black}{\figurename~\thefigure}}
\makeatother
    \includegraphics[width=0.6\textwidth]{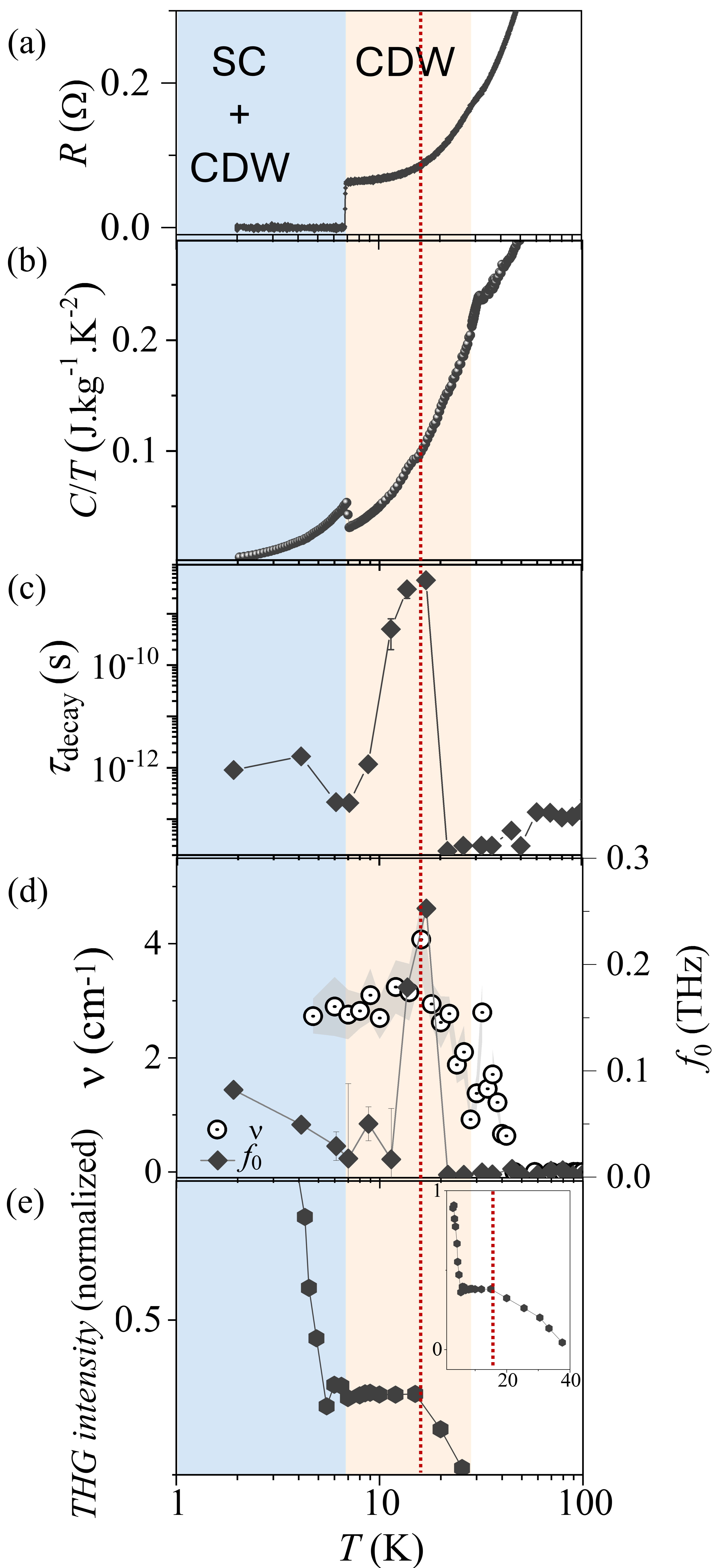} 
    \caption{{(a)-(e) plot the electrical resistance ($R$) the specific heat ($C/T$), electron relaxation time ($\tau_{decay}$), overdamped oscillation frequency ($f_0$), and Fano coupling parameter ($\nu$), and THz-THG area, respectively. The inset of (e) shows the linear full-scale plot of the THG intensity.}}
    \label{specific heat}
\end{figure*}

Fig.~\ref{specific heat}(b) shows the specific heat measured in a different batch of 2\textit{H}-NbSe$_2$ crystals. Ambient-pressure electrical transport and heat-capacity measurements were performed using a Quantum Design PPMS, with temperatures (T) down to 1.8 K. The specific heat data were measured using a relaxation technique.
$C/T$ shows clear change at $T_c$ and $T_{CDW}$, as well as a small kink close to 14~K, indicated by the black arrow. This is in line with the onset of a new dynamics setting in the system and the quasi-divergence of $\tau_{decay}$ at this temperature.

\textcolor{black}{We use high-field multicycle 0.3 THz pulses focused to spot size of the order of 1 mm FWHM on the sample and an experimental geometry, as discussed in \cite{PhysRevB.108.L100504}, to filter for the third harmonic generation (THG) signal at 0.9 THz. 
}
\textcolor{black}{Fig.~\ref{specific heat}(e) plots the THz-THG intensity as a function of temperature. We observe an order-parameter-like behaviour setting at Tcdw. We further note a kink-like feature at 17 K, below which the response saturates. This robustness of the order parameter behaviour of CDW below this temperature is consistent with a strong coupling to the interlayer shear phonons and formation of a new coherent state.
This feature was already observed in \cite{PhysRevB.108.L100504} but not discussed. Here, we reproduce the feature, highlighting the flattening behavior of the order parameter below 17 K.}

\textcolor{black}{Fig.~\ref{specific heat} (a)-(e) plot the electrical resistance, specific heat and the dynamical parameters extracted from the ultrafast response, Raman spectroscopy, and THz-THG measurements. We observe noticeable changes at 17 K (indicated by the red dotted line) across all observables in independent experiments of specific heat measurements, ultrafast, Raman and THz spectroscopy performed in varying batches of samples. These are all consistent with the onset of a collective phase of CDW puddles driven by the Fano-coupled ph-CDW hybrid. }
\clearpage
\newpage

\end{document}